# Direct detonation initiation in hydrogen/air mixture: effects of compositional gradient and hotspot condition

*Xiongbin Jia[1,2], Yong Xu[2], Hongtao Zheng[1], and Huangwei Zhang[2*]*

[1] *College of Power and Energy Engineering, Harbin Engineering University, Harbin, 150001, China*
[2] *Department of Mechanical Engineering, National University of Singapore, 9 Engineering Drive 1, Singapore 117576, Singapore*

## Abstract

Two-dimensional simulations are conducted to investigate the direct initiation of cylindrical detonation in hydrogen/air mixtures with detailed chemistry. The effects of hotspot condition and mixture composition gradient on detonation initiation are studied. Different hotspot pressure and composition are first considered in the uniform mixture. It is found that detonation initiation fails for low hotspot pressures and supercritical regime dominates with high hotspot pressures. Detonation is directly initiated from the reactive hotspot, whilst it is ignited somewhere beyond the nonreactive hotspots. Two cell diverging patterns (i.e., abrupt and gradual) are identified and the detailed mechanisms are analyzed. Moreover, cell coalescence occurs if many irregular cells are generated initially, which promotes the local cell growing. We also consider nonuniform detonable mixtures. The results show that the initiated detonation experiences self-sustaining propagation, highly unstable propagation, and extinction in mixtures with a linearly decreasing equivalence ratio along the radial direction respectively, i.e., 1→0.9, 1→0.5 and 1→0. Moreover, the hydrodynamic structure analysis shows that, for the self-sustaining detonations, the hydrodynamic thickness increases at the overdriven stage, decreases as the cells are generated, and eventually become almost constant at the cell diverging stage, within which the sonic plane shows a "sawtooth" pattern. However, in the detonation extinction cases, the hydrodynamic thickness continuously increases, and no "sawtooth" sonic plane can be observed.

[*] Corresponding author. Email: Huangwei.zhang@nus.edu.sg.

## 1. Introduction

Detonation propulsion, e.g., rotating detonation engine, has great potential because of high thermal efficiency and simple structure. Efficient detonation initiation is critical to materialize this technology with compact engine structure and reliable operation. Typically, detonative combustion can be ignited by indirect and direct initiation. For the latter, a detonation can be initiated when the deposited energy is sufficiently high (Body 1997), e.g., through spark (Matsui & Lee 1976) or detonating cord (Higgins, Radulescu & Lee 1998). However, due to extremely short space and time scales, detailed detonation initiation and development are difficult to be captured experimentally (Radulescu *et al.* 2003), and hence our understanding about the underlying mechanism is still rather limited.

It is well known that a critical energy $E_c$ exists to directly initiate a detonation wave in a detonable mixture (Zhang & Bai 2014). Depending on the deposited energy $E_s$, three regimes are identified: supercritical ($E_s > E_c$), critical ($E_s \approx E_c$), and subcritical ($E_s < E_c$) regimes (Ng & Lee 2003). Zel'dovich (1956) propose a theoretical model to determine the critical energy $E_c$, where $E_c$ varies exponentially with the induction length. However, due to a series of simplifications involved, their criterion is applicable for stable detonation, in which the induction length is relatively small. After that, several prediction models are developed, e.g., by Lee, Knystautas & Guirao. (1982), Zhang, Ng & Lee (2012), and Ng (2005), in which an average delay in ignition is applied. The critical energy predicted by these models is in good agreement with the experimental data since well estimated detonation parameters are incorporated, such as detonation cell size and critical tube diameter. However, the detonation front is intrinsically unstable and exhibits a complex triple-point structure, which plays a key role in the direct detonation initiation (Shen & Parsani 2017).

Taking detonation curvature and unsteadiness into account, Kasimov and Stewart (2004) establish a prediction model named as $\overline{D} - D - \kappa$ ($\overline{D}$ is detonation wave acceleration, $D$ the detonation speed, and $\kappa$ the curvature), using single-step chemistry. Their model is improved by Soury and Mazaheri (2009), who incorporate detailed chemical kinetics and predict better relations between $E_c$ and the equivalence ratio. However, the model works for limited mixture composition due to the complex chemical reaction process and multi-dimensional effects during direct initiation (Zhang & Bai 2014). Furthermore, some details, including time-dependent detonation structure variations, and its effects of unburned pockets on the direct initiation process, cannot be elucidated by these models.

Different from planar detonations, the curvature plays an important role in cylindrical and spherical detonations. He (1996), Eckett, Quirk & Shepherd (2000), Watt & Sharpe (2004; 2005), and Han *et al.* (2017) demonstrate the destabilizing effect of global curvature on detonation waves, and larger curvature would aggravate these effects. For instance, He (1996) find that a maximum curvature is defined by the the nonlinear curvature effect, beyond which a self-sustaining detonation cannot be obtained. Eckett, Quirk & Shepherd (2000) point out that the unsteadiness in the induction zone is responsible for failure of detonation initiation. Watt & Sharpe (2004; 2005) show that the pulsation amplitude arising from the curvature varies with the radius with which the detonation is first generated. Considering cellular stability, Han *et al.* (2017) find that the detonation structure evolves following three stages, i.e., no cell, growing cells, and diverging cells. They also analyze the weakening effect of unburned pockets on average detonation speed as the detonation cell increases (hence curvature decreases).

Most previous work on direct initiation is focused on one-dimensional problems where only longitudinal pulsating instability is incorporated (Ng & Lee 2003; He 1996; Eckett, Quirk & Shepherd 2000; Watt & Sharpe 2004; Watt & Sharpe 2005; Han *et al.* 2017; Qi & Chen 2017). Nonetheless, in realistic situations, multi-dimensional cellular instability should be considered. Shen & Parsani (2017, 2019) study the effects of multi-dimensional instabilities on the direct initiation by comparing the phenomena from one- and two-dimensional simulations. Their results show that the one-dimensional configuration becomes invalid for unstable detonations. They also emphasize the important role of strong transverse waves from multi-dimensional instabilities in the failure and initiation process of detonation. Moreover, Han *et al.* (2018) examine the effects of activation energies of chemical kinetics on detonation initiation. They find that the continuous propagation of cellular detonation with higher activation energies exhibit a stronger dependence on regeneration of the transverse wave. Furthermore,

Jiang *et al.* (2009) identify four mechanisms of the cell diverging in cylindrical detonation expansion based on two-dimensional simulations. This provides a deeper understanding on the relation between flow instability and generation/diminishing of transverse waves in different patterns. Besides, Asahara *et al.* (2012) further shows the detailed Mach configuration and generation of sub-transverse waves during the cell diverging process.

Past numerical work on direct detonation initiation is mostly concentrated on uniform mixture. To the best of our knowledge, research on direct initiation of a detonation in non-uniform mixtures is still lacking. Furthermore, the effects of hotspot properties (e.g., reactive, or nonreactive) on the mechanisms of detonation initiation have not been well understood. In the current study, we aim to examine the effects of hotspot properties and mixtures composition gradient on direct detonation initiation. Two-dimensional simulations with detailed chemical mechanism will be conducted. The manuscript is organized as follows: section 2 presents the governing equation and numerical method, whilst the results and discussion are detailed in sections 3 and 4, respectively, followed by section 5 with conclusions.

## 2. Mathematical and physical models

### *2.1. Numerical method*

The Naiver-Stokes equations of mass, momentum, energy, and species mass fractions are solved for compressible reacting flows, with the solver *RYrhoCentralFoam* (Xu *et al.* 2021). The accuracies of *RYrhoCentralFoam* in detonation simulations have been extensively validated (Huang *et al.* 2021), and it has been used for various detonation problems (Huang, Cleary & Zhang 2020; Huang & Zhang 2020; Xu, Zhao & Zhang 2021). Second-order implicit backward method is employed for temporal discretization, and the time step is $1\times10^{-9}$ s. A Riemann-solver-free MUSCL scheme (Kurganov, Noelle & Petrova 2001) with van Leer limiter is employed to calculate the convective fluxes in the momentum equations. The total variation diminishing scheme is used for the convection terms in energy and species equations, whilst a second-order central differencing scheme is adopted for the diffusion terms in equations of momentum, energy, and species mass fractions (Huang *et al.* 2021; Greenshields *et al.* 2010). A detailed hydrogen mechanism is applied, with 13 species and 27 reactions (Burke *et al.* 2012). The chemical source term is integrated with an implicit Euler method.

### *2.2. Physical problem and numerical implementation*

The expanding cylindrical detonation has the intrinsic characteristic of cellular instability, which plays an important role in initiation of transverse waves (Shen & Parsani 2017; Han *et al.* 2017). In this work, two-dimensional simulations (see Fig. 1) are conducted to capture the detonation frontal instability and dynamic behaviors. Due to the geometrical symmetry, a quarter area of the domain is simulated, and the domain is 0.5×0.5 $m^2$ (see Fig. 1). The *x* and *y* axes are aligned with the symmetry boundaries, and the radius is $R = \sqrt{(x^2 + y^2)}$. For two outlets, wave-transmissive condition is enforced for the pressure, whereas zero-gradient condition for all rest quantities.

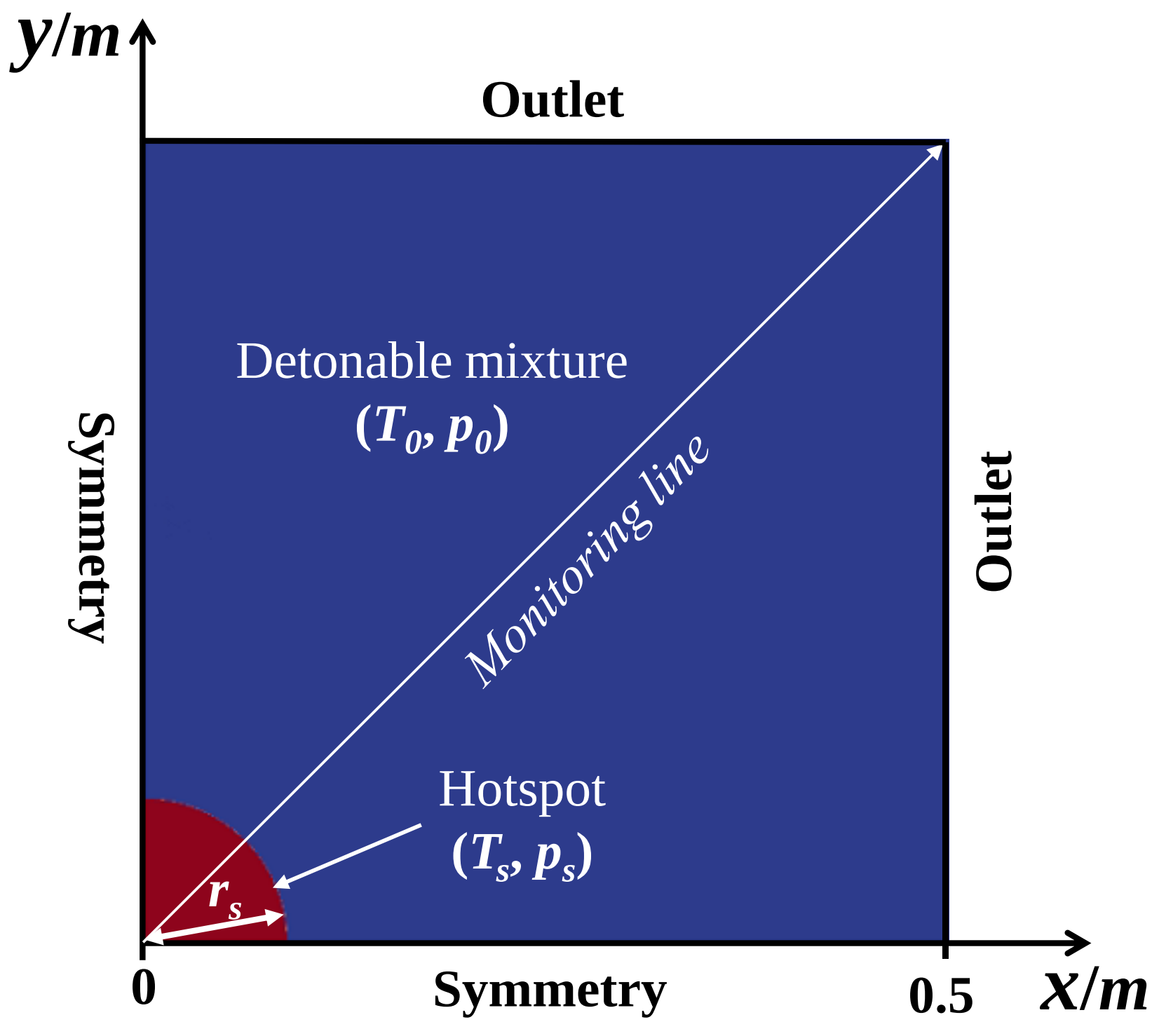

Figure 1: Schematic of the computational domain and boundary condition. Hotspot size not to scale.

The domain consists of two parts, as illustrated in Fig. 1. The first part is the circular hotspot with high temperature and pressure, ($T_s$, $p_s$), to mimic a localized ignition, e.g., resulting from additional energy deposition or shock focusing. The radius is fixed to be $r_s$ = 0.02 m in the simulations. Varying the hotspot size may affect the detonation initiation (Lee & Ramamurthi 1976), but we will not study it in this paper. Both non-reactive (the hotspot composition is air) and reactive ($H_2$+$O_2$ or $H_2$+air) hotspots are considered.

The second part, beyond the hotspot, is filled with quiescent detonable gas, i.e., $H_2$+$O_2$+$N_2$ mixtures. The initial pressure is $p_0$ = 20265 Pa and the initial temperature is $T_0$ = 300 K. Both uniform and varying composition of the detonable mixture will be studied. For the former, the composition of the gaseous mixture is $H_2$:$O_2$:$N_2$ = 0.0282:0.2255:0.7463 by mass. For the latter, linear change of the equivalence ratio along the radial direction will be considered to examine its effects on detonation initiation and subsequent development.

The uniform 62.5 μm Cartesian cells are adopted to discretize the domain in Fig. 1, and the total mesh number is about 64 million. The half-reaction length from the theoretical ZND structure is approximately $l_{1/2}$ = 1 mm, calculated by the Shock and Detonation Toolbox (Shepherd 2021), and hence the foregoing mesh size corresponds to about 16 pts/$l_{1/2}$ for a CJ detonation. The mesh sensitivity test is shown in Section A of the supplementary document, and the results show that mesh convergence can be obtained when the mesh resolution of 16 pts/$l_{1/2}$ is employed.

Table 1: Information of simulated cases

| Case | Effects | | Hotspot properties | Detonable mixture equivalence ratio | Regime |
|---|---|---|---|---|---|
| A | Hotspot properties | pressure | 250$P_0$; 2500 K; air | 1 | Critical |
| B | | | 200$P_0$; 2500 K; air | | Critical |
| C | | | 150$P_0$; 2500 K; air | | Sub-critical |
| D | | | 100$P_0$; 2500 K; air | | Sub-critical |
| E | | composition | 100$P_0$; 2500 K; $H_2$+air | | Super-critical |
| F | | | 100$P_0$; 2500 K; $H_2$+$O_2$ | | Super-critical |
| G | Mixture composition gradients | | 100$P_0$; 2500 K; $H_2$+$O_2$ | 1→0.9 | Super-critical |
| H | | | | 1→0.5 | |
| I | | | | 1→0 | |

*2.3. Simulation case*

Parametric studies are performed and nine cases, i.e., A-I, are selected for discussion in this paper (see details in Table 1). Specifically, case A-F have different hotspot parameters, whereas case F-I different composition gradients in the detonable gas. When the air is filled in the hotspot, critical regime dominates for relatively high $p_0$ (e.g., A and B), whereas critical regime for low $p_0$ (e.g., C and D). Moreover, super-critical regime is observed with reactive hotspots (e.g., E and F). To study the mixture composition gradients, reactive hotspot with $H_2$+$O_2$ (same with F) is selected to ensure successful initiation. Three equivalence ratio gradients are considered. Specifically, the equivalence ratio in the vicinity of the hotspot (the radius $R$ = 0.02 m) is fixed to be 1, which decreases linearly to a certain value (e.g., 0.5, see Table 1) at the outer edge ($R$ = 0.5 m). For easy reference, we use an arrow to indicate the radial ER change, e.g., 1→0.9 in case G.

## 3. Results

*3.1. Effects of hotspot properties*

In this section, we will study the effects of hotspot pressure and gas composition on detonation initiation and development in a uniform detonable gas.

*3.1.1 Hotspot pressure effects*

Figure 2 shows the leading shock speed versus the radius with different hotspot pressures, i.e., $p_s$ = 250$p_0$, 200$p_0$, 150$p_0$, 100$p_0$. They are case A-D, and the hotspot is filled with air. The shock speed is estimated along the radial monitoring line (see Fig. 1). The detonation is successfully initiated only when $p_s \geq 200p_0$. Under relatively low hotspot pressures, the shock from the hotspot decelerates quickly and the detonation initiation fails. For instance, with $p_s$ = 150$p_0$, the shock speed increases from 0.03 to 0.035 m after an initial drop (see the inset of Fig. 2). This is because intense reactions are triggered to release energy intensifying the shock. However, due to fast shock decay, the reaction front (RF) fails to coherently couple with the leading shock front (SF). Therefore, the latter degrades to a blast wave with a speed of around $0.3V_{CJ}$, corresponding to the sub-critical regime (Ng & Lee 2003).

With $p_s$ = $200p_0$ and $250p_0$, the shock from the ignition spot steeps into an overdriven detonation, due to the strengthening effects from the shocked mixture. The detonation gradually decays to a freely propagating detonation around 0.1 m. This can be categorized into the critical regime. The average propagation speed is slightly lower than the CJ speed, which is caused by the curvature effects (Ng & Lee 2003).

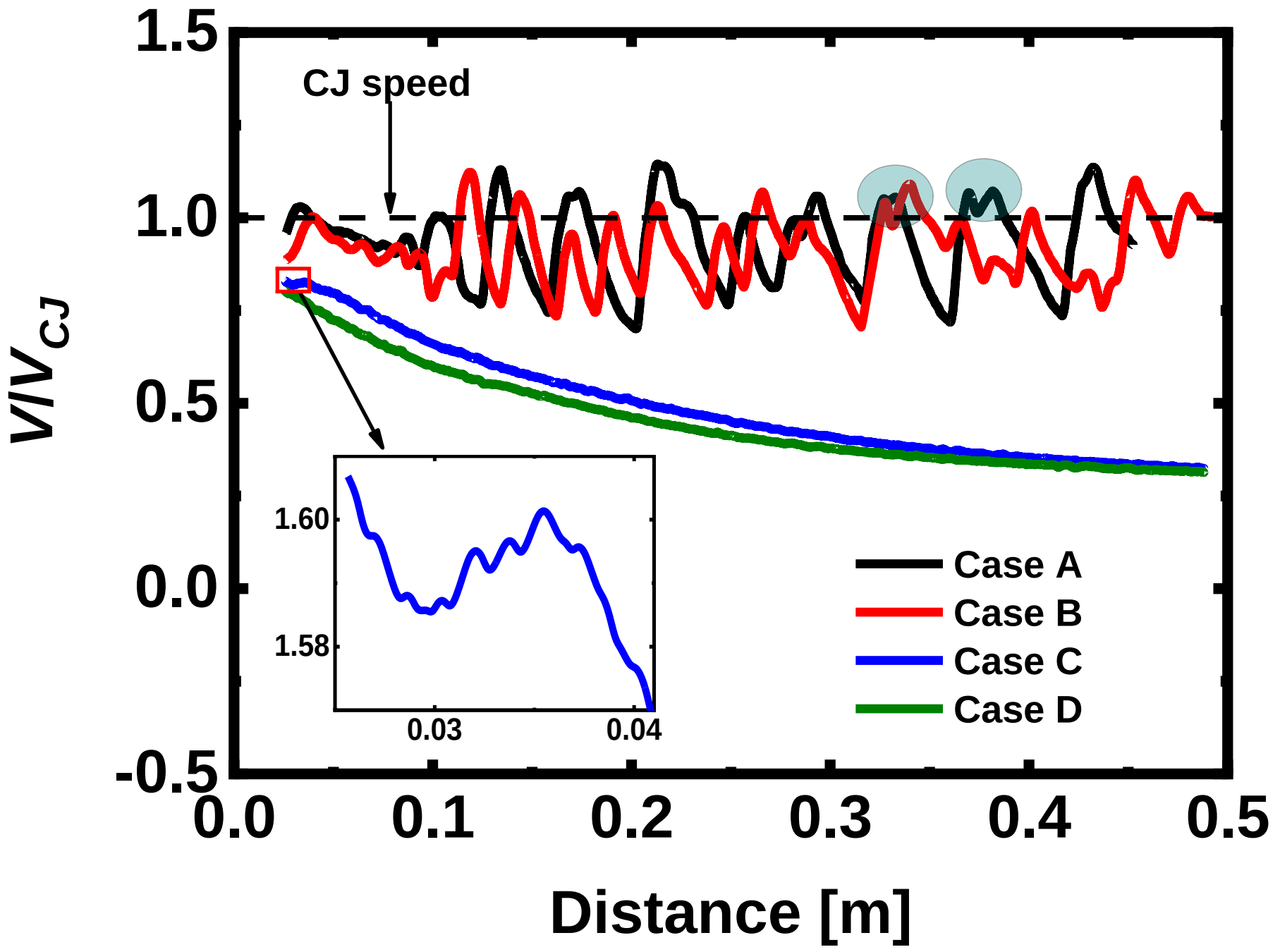

Figure 2: Change of the leading shock speed with radial distance when different hotspot pressures are considered (case A-D).

Figure 3 shows the detonation cell evolutions recorded from the peak pressure trajectories in case A and B. For cylindrical detonations, the cell size λ is defined from the azimuthal direction (roughly perpendicular to the detonation propagation direction) (Lee 1984). For both cases, the detonation cell experiences three stages as the front curvature decreases: no cell (I), growing cells (II), and diverging cells (III), as annotated in Fig. 3. This is consistent with the observation by Han *et al* (2017). However, the detonation cell growing is not quantified in their study, and the detailed diverging process and mechanism still remain to be revealed. In diverging cells stage, new cells are generated from the enlarged cells. As such, some small fluctuations of the shock speed along the monitoring line are superimposed on the original periodic fluctuations, leading to double peaks, see the circles in Fig. 2. Besides, the detonation cell evolution before the cell diverging stage (III) is featured by a series of cell-family. The results show that the number of detonation cell keeps almost constant as the cells grow. As such, the detonation cell increases linearly with the distance. Besides, higher hotspot pressure generates more cell-family numbers, corresponding to globally smaller cell size. The calculated cell-family numbers in case A and B are about 17 and 12, respectively. This is justifiable because large hotspot pressure results in high overdrive degree of the newly generated detonation wave. Besides, at $R$ > 3.5 m, the cell begins to diverge in both cases. The diverging happens only in locally larger cells in case A (see red circles), but in larger domain in case B with greater growth rate.

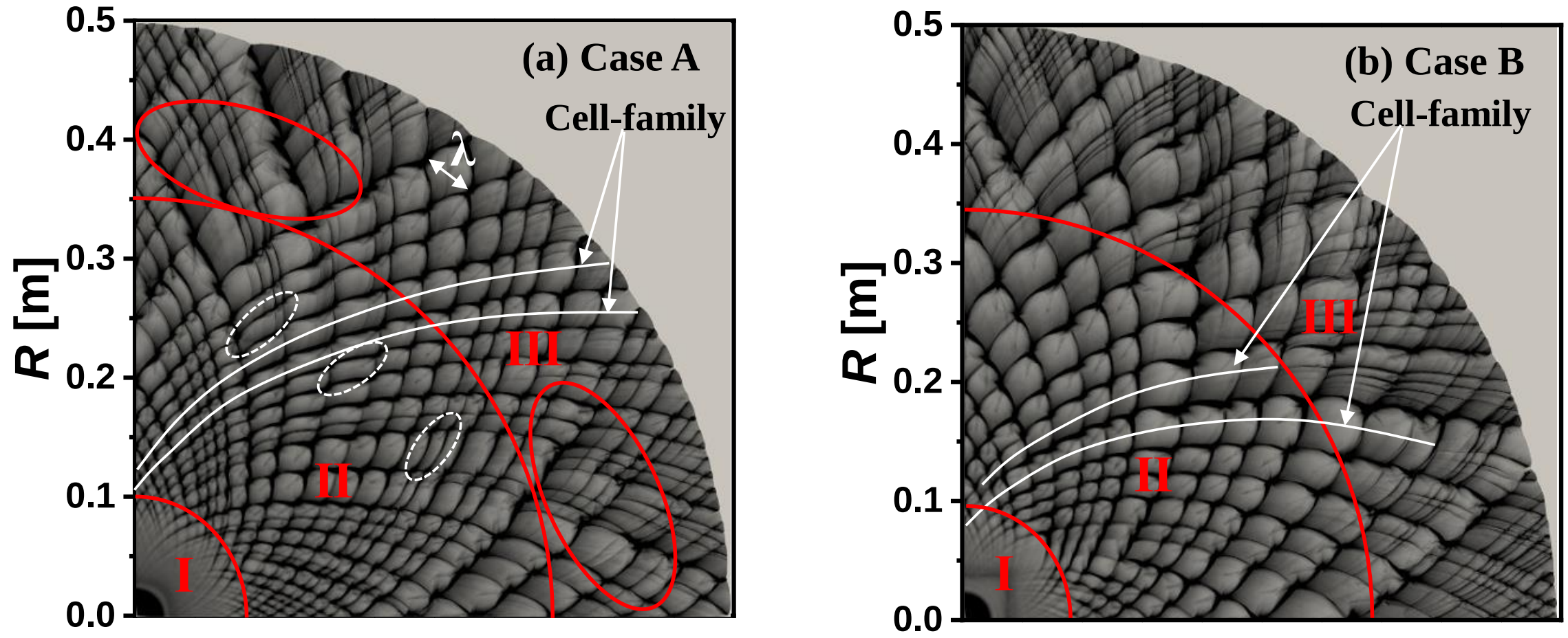

Figure 3: Detonation cell evolutions with different hotspot pressures: (a) case A, $p_s = 250p_0$; (b) case B, $p_s = 200p_0$. Stage I: no cell; II: cell growing; III: cell diverging.

*3.1.2 Hotspot composition effects*

Here we will further examine the influences of hotspot composition on detonation initiation and development. Figure 4 shows the shock speed evolutions when two reactive hotspots are considered, i.e., stoichiometric $H_2$+air and $H_2+O_2$ mixtures. They are case E and F, respectively. The result with air hotspot (i.e., case D) is included for comparison. The hotspot pressure is $100p_0$. Different from the observation in Section 3.1.1, an overdriven detonation is directly initiated and then decays to CJ detonation in both case E and F, which correspond to the super-critical regime (Ng & Lee 2003). As the overdrive degree decays, the periodic fluctuations of the shock speed occur earlier in case E, indicating an earlier onset of detonation cellularization. Furthermore, the speed fluctuation in case E exhibits more irregularity, especially at larger radii (hence smaller curvature).

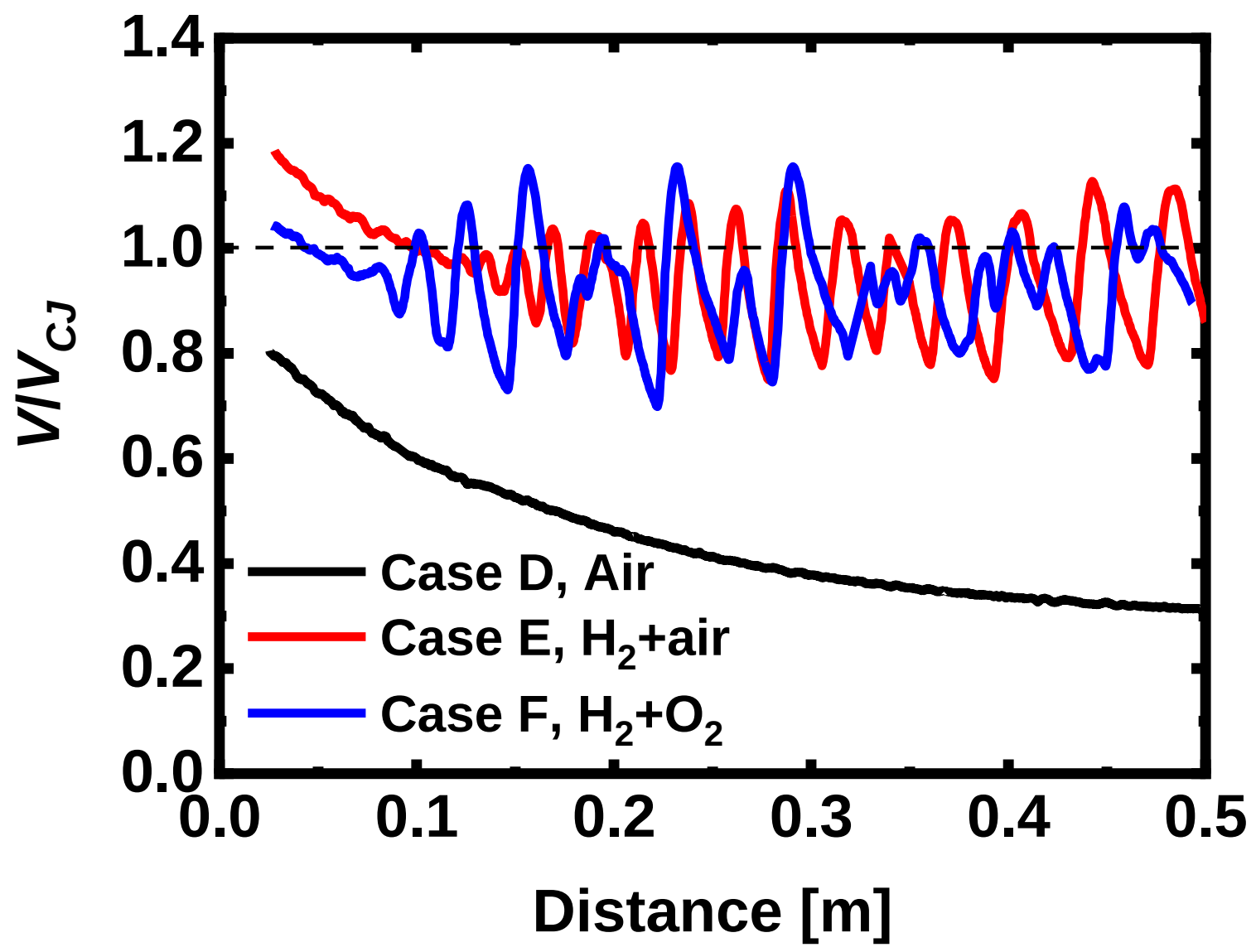

Figure 4: Change of the leading shock speed as a function of radial distance in case D-F. $p_s$ = $100p_0$.

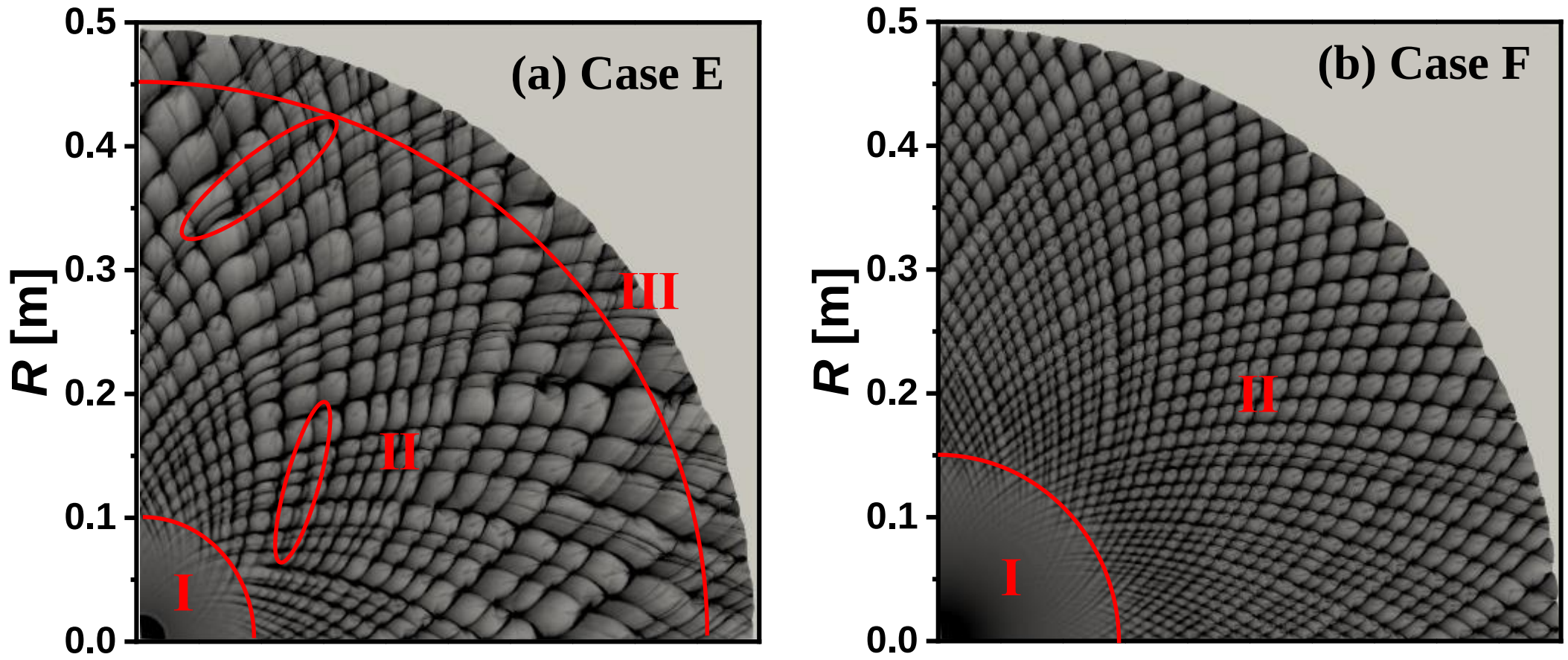

Figure 5: Detonation cell evolutions with different hotspot compositions: (a) case E: stoichiometric $H_2$+air mixture; (b) case F: stoichiometric $H_2$+$O_2$ mixture. Stage I: no cell; II: cell growing; III: cell diverging.

Figure 5 shows the cell evolutions in case E and F. In general, the cells in case F ($H_2$+$O_2$) are smaller and more uniform. Only two stages appear, i.e., no cell (I) and growing cell (II). Even at the maximum radius in our simulation, the cell is still too small to diverge. Furthermore, in case E ($H_2$+air), some relatively small cells merge to the adjacent larger ones at the second stage, see the red circles in Fig. 5(a). This phenomenon will be further discussed in Section 4.2. Moreover, cell diverging occurs at $R$ = 0.45-0.5 m. The cell inhomogeneity and diverging behaviour lead to irregular shock speed fluctuations in Fig. 4. The cell-family number in E and F are 18 and 31, respectively. Consequently, the cell in F grows more slowly and its maximum cell size reaches about 27 mm, slightly smaller than the theoretical value, 28.9 mm (Ng, Ju & Lee 2007). In this sense, the detonation can still propagate in a self-sustaining fashion without new cell generation. Combined with Fig. 3, it can be found that there exists a range within which cell diverging is more likely to take place, and in section 4.2 we will further study this range.

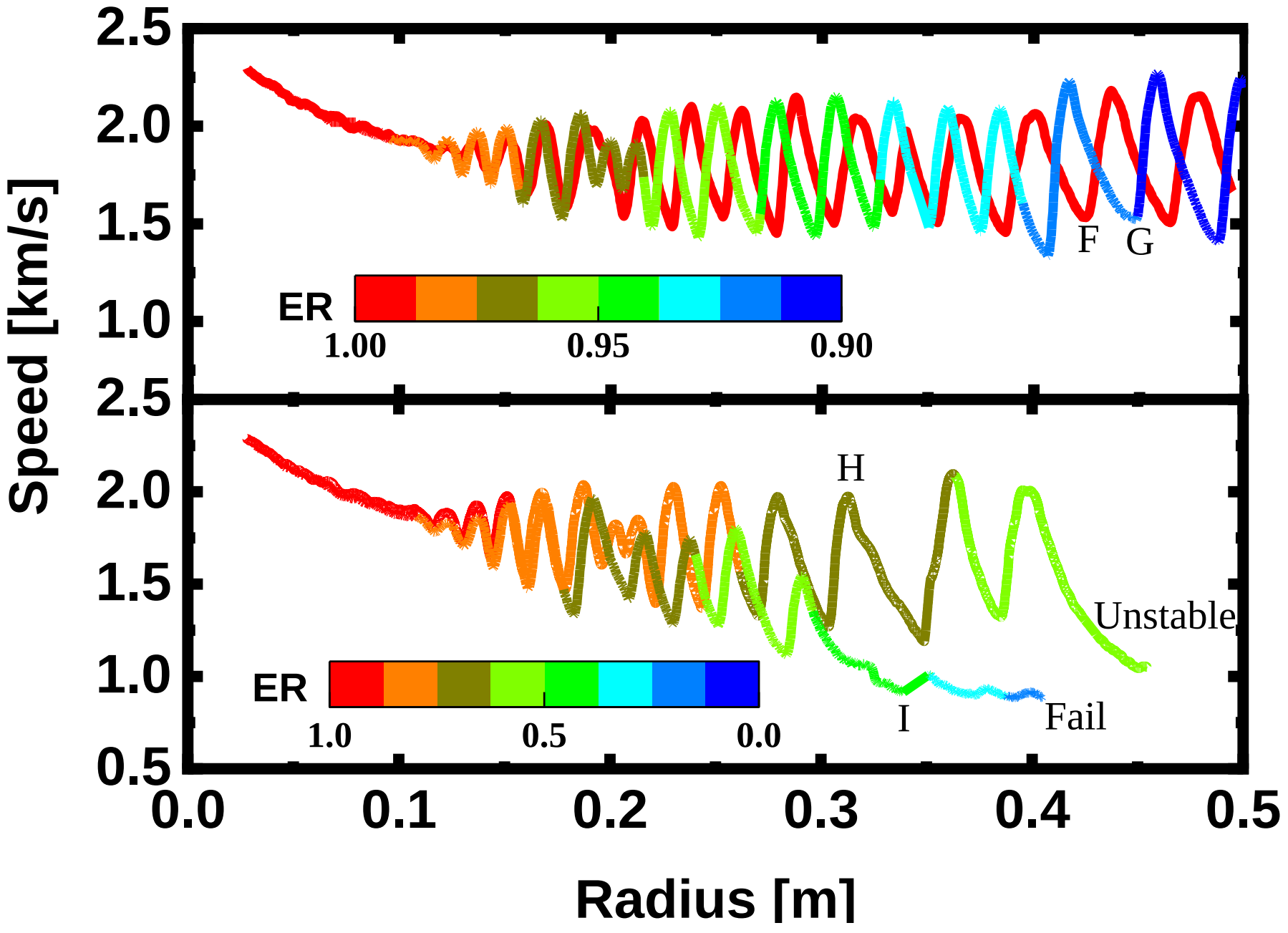

Figure 6: Change of the leading shock speed as a function of radial distance in case F-I. The profiles are coloured by the local equivalence ratio.

### 3.2. Effects of composition gradient in the detonable mixture

In this section, we will study detonation initiation in $H_2$+air mixtures with spatially varying equivalence ratio. Specifically, the ER in the vicinity of the hotspot is 1, and then decreases linearly to 0.9, 0.5 and 0 at $R$ = 0.5 m in case G, H, and I, respectively. A hotspot ($100p_0$, 2,500 K) with stoichiometric $H_2$+$O_2$ mixture is employed, same as that in case F.

Figure 6 shows the variations of the leading shock speed in case F-H along the monitoring line. Since the reactive hotspot ($H_2$+$O_2$) is employed, overdriven detonations are directly triggered by it in all cases. Generally, their shock speeds are close before 0.1 m, due to the near-stoichiometric ERs. Beyond that, significant differences appear when the multi-headed detonations start to develop, featured by the various speed fluctuations. Among them, continuous detonation propagation happens only in case F (uniform, φ: 1→1) and G (φ: 1→0.9). The average shock speed during the cellular detonation stage ($R$ = 0.12-0.5 m) in G is 1,756 m/s, slightly lower than that in F (1,768 m/s) since the mixture reactivity decreases in the former case (see Section B of the supplementary document). Moreover, in case H and I, the overdriven detonation gradually decouples when it runs outwardly. Specifically, in case H (φ: 1→0.5), when the shock propagates across $R$ = 0.3 m (the local ER is φ = 0.71), the period of speed fluctuation increases significantly, indicating more unstable detonation front. In case I (φ: 1→0), the detonation wave quickly decouples across $R$ = 0.3 m (φ = 0.42) with a shock speed below 900 m/s.

Figure 7 shows the detonation cell evolutions in case F-I. Almost uniform cells appear at about 0.15 m and grows until 0.5 m in both F and G, except for locally larger cells, as annotated by the red circles in Figs. 7(a) and 7(b). This phenomenon is attributed to the cell-family differences (i.e., 31 for F, 27 for G) and cell coalescence (see the inset in Fig. 7), which are caused by the drop of mixture reactivity with the decrease of ER in case G (see Section B of the supplementary document), especially at larger radii.

Increased composition gradient in case H and I lead to remarkable change of detonation cell distribution, as shown in Figs. 7(c) and 7(d). In case H (φ: 1→0.5), the cells grow steadily from $R$ = 0.15 to 0.3 m and remain diamond shaped. Beyond that, they become irregular, and some cells grow faster and then merge with the adjacent smaller cells. As the detonation propagates across $R$ = 0.25 m (φ = 0.75), irregular cell growing appears as the reactivity of mixtures drops significantly (see Section B of the supplementary document). Therefore, apart from the curvature decrease which leads to the increase of the detonation cell size, the cell coalescence induced by the ER variation plays a more important role in the oversized cell generation. This oversized cell further causes the local detonation quenching. At around $R$ = 0.5 m, the detonation becomes very unstable, and detonative combustion is even quenched at most of the front. In case I (φ: 1→0), the detonation cells appear at around $R$ = 0.13 m and slightly grows until 0.3 m, like the rest cases in Fig. 7. However, detonation extinction happens beyond $R$ = 0.3 m (φ = 0.42), with quickly faded peak pressure trajectories in Fig. 7(d).

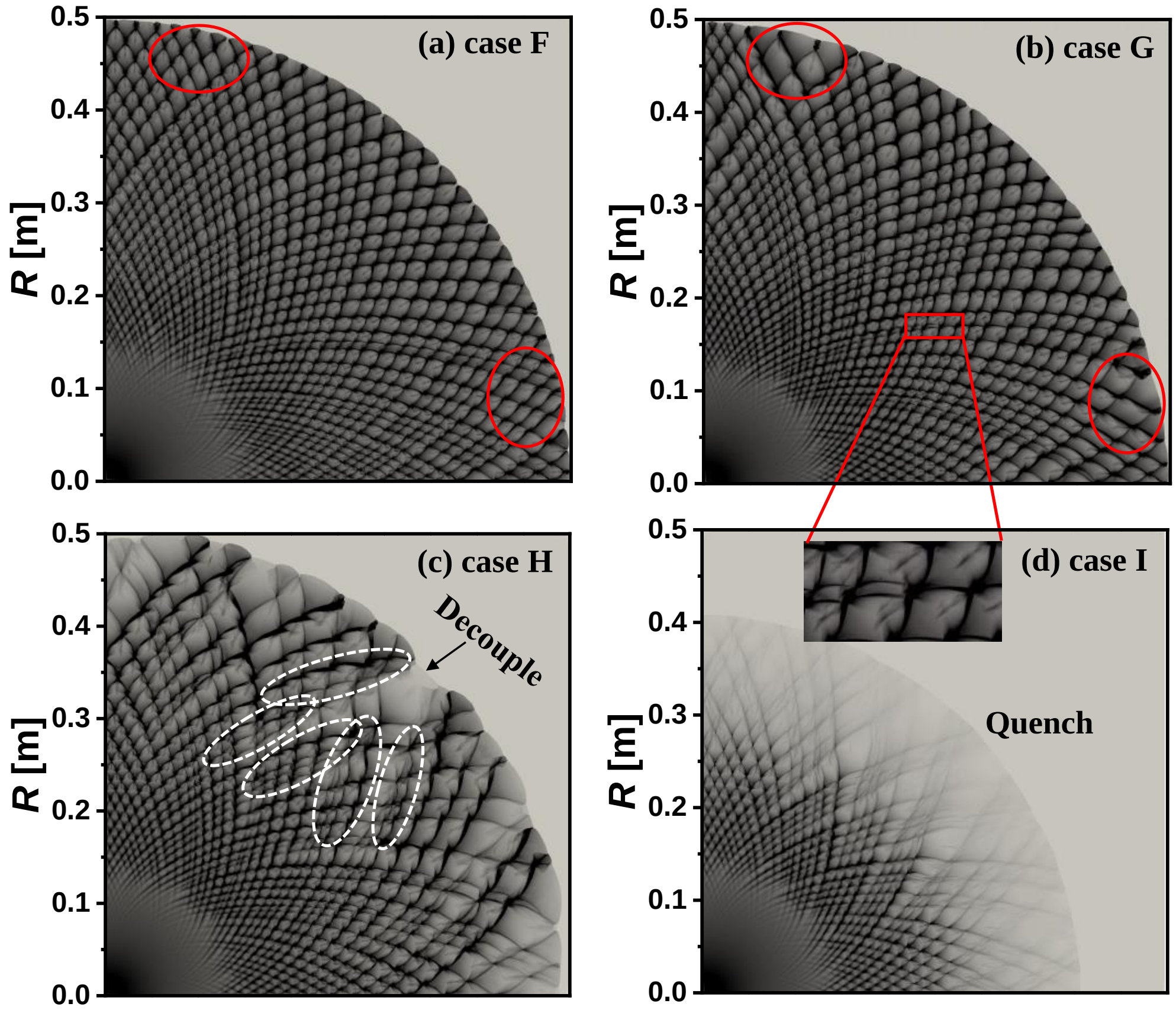

Fgure 7: Detonation cell evolution with different mixture composition gradients in case F-I.

Figure 8 shows the time evolutions of the shock speed and SF/RF position along the monitoring line in case I. Three stages can be identified: overdriven detonation (0-0.05 ms), cellular detonation (0.5-0.13 ms), and detonation quenching (0.13-0.2 ms). At the cellular detonation stage, the shock speed experiences periodic variations with an increasing time interval. The inset in Fig. 8(a) provides the change of SF/RF in the cellular detonation stage. The red arrows indicate the abrupt acceleration of the RF, whereas the black ones the generation of the MS. From Fig. 8(b), the speeds of both fronts fluctuate after 0.05 ms and the periodic variations of the SF speed are delayed relative to that of the RF, manifesting an intrinsic characteristic of unstable cellular detonation. Since 0.13 ms, the RF speed continuously decreases and is lower than that of the SF, which indicates the decoupling of SF and RF.

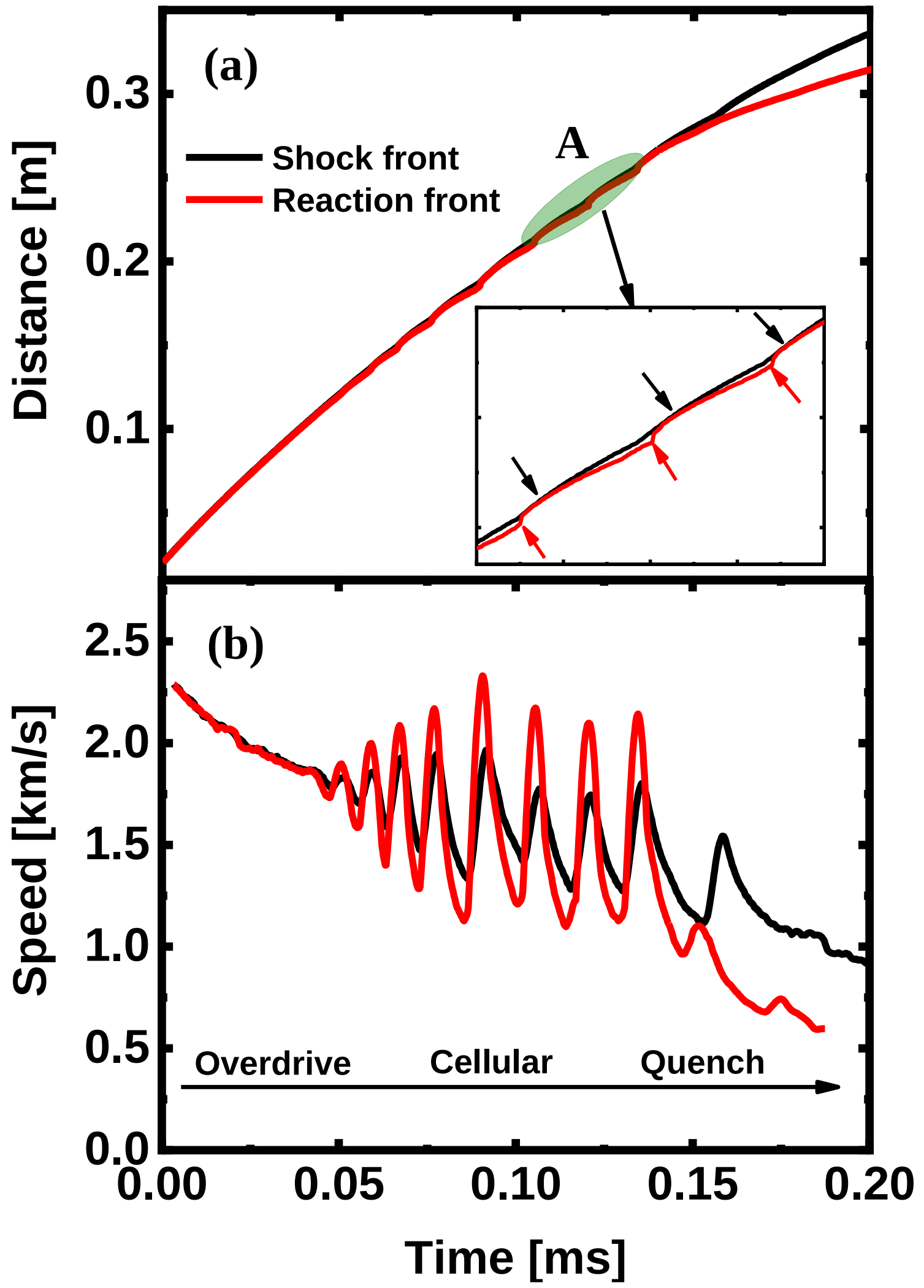

Figure 8: Time history of (a) SF and RF position and (b) propagating speed along the monitoring line in case I.

Figure 9 shows the distributions of temperature and hydrogen mass fraction at different instants during the detonation extinction process. From 126-134 μs, the Mach stem, MS1, gradually attenuates and evolves to an incident shock, IS1, with the unburned pocket 1 (U1) left behind, which weakens the SF intensity. Meanwhile, a longer induction zone is formed behind IS1. Two detonation bubbles are generated from the local explosion induced by shock focusing (Lee 2008) (see the red circle at 126 μs), which evolves into the MS2 and MS3 (see 134 μs). Due to the increased induction length, the new ISs collide with each other, leading to weaker focusing energy (see 142 μs). Therefore, another larger unburned pocket (U2) is generated, which further reduces the local heat release. Another new IS4 develops from the focusing and propagates outwardly; however, the reaction cannot be triggered by this weak focusing, e.g., at 150 μs. Consequently, the detonation quenches and further degenerates into an inert shock wave at 162 μs. Meanwhile, when the RF and SF are fully decoupled (at 162 μs), considerable unburned $H_2$ exists behind the shock, and the low temperature can be found between the SF and RF.

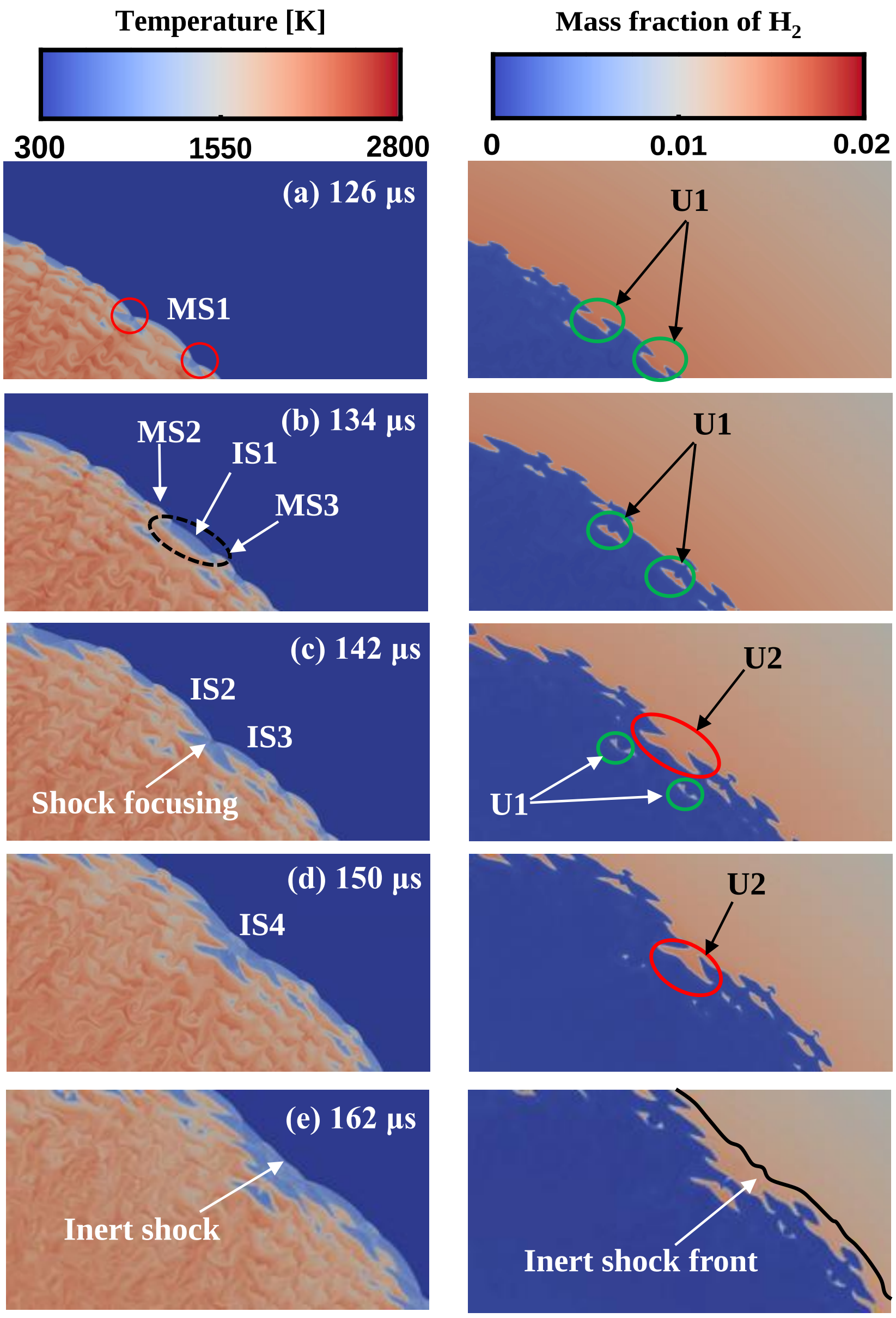

Figure 9: Time sequence of temperature and hydrogen mass fraction in case I.

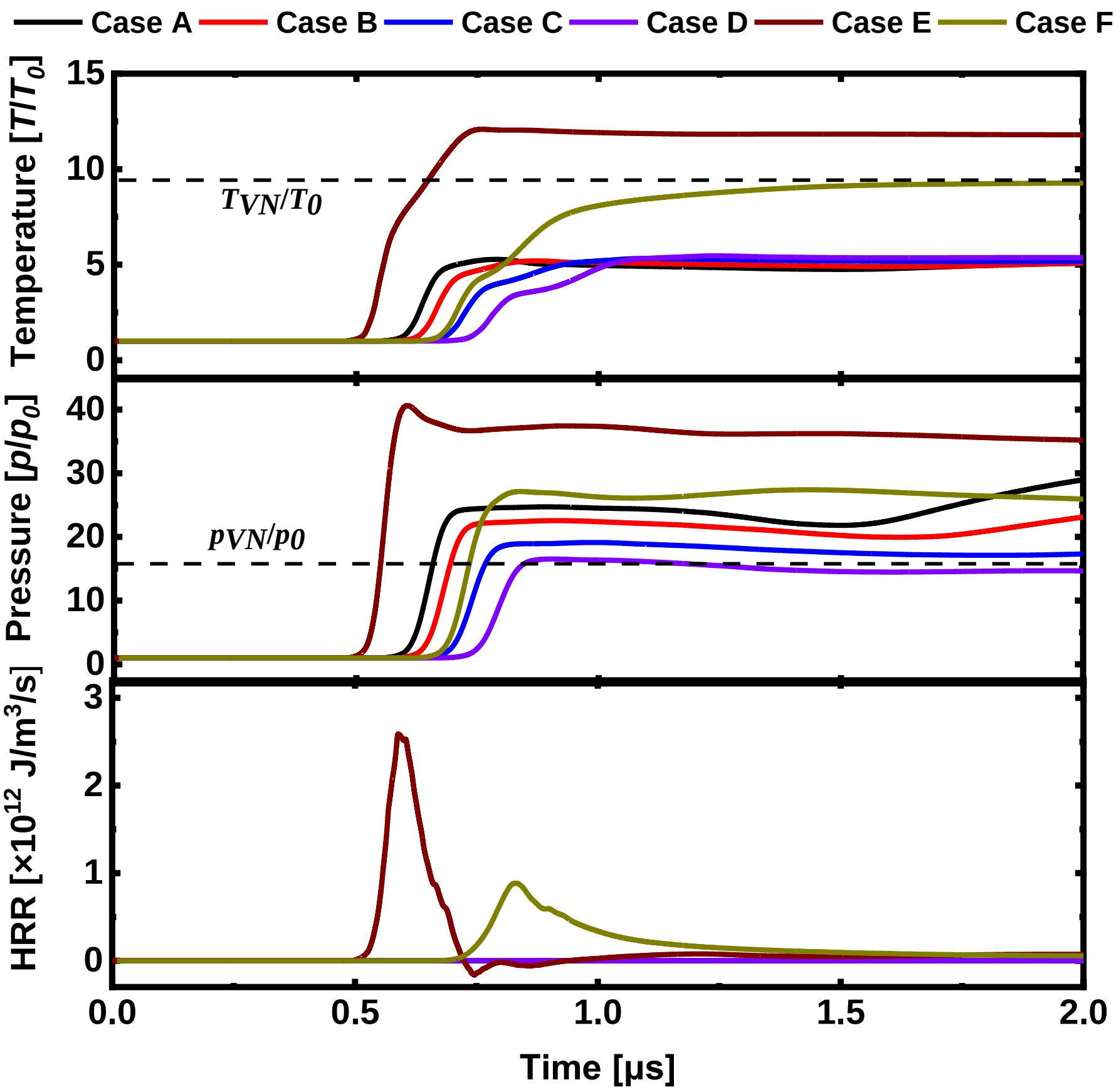

Figure 10: Evolutions of (a) temperature, (b) pressure, and (c) HRR from the probe ($R$ = 21 mm). The von Neumann states are from the detonable gas condition ($H_2$+air mixture).

## 4. Discussion

### *4.1. Hotspot evolution and their relevance to detonation initiation*

Hotspot evolution and their relevance to detonation initiation will be discussed here based on case A-F. Figure 10 shows the time history of temperature, pressure, and HRR, from a probe of $R$ = 21 mm, i.e., 1 mm off the hotspot vicinity along the monitoring line. It is shown that intense chemical reactions take place in two reactive hotspots (i.e., case E and F) with the maximum pressure and temperature higher than the corresponding von Neumann values. This indicates that the detonations are initiated directly from the hotspot. Note that the detonation in case E manifests a higher overdrive degree ($f$ = 1.41) than that in F ($f$ = 1.09) due to higher oxygen concentration (Short & Stewart 1999). As the overdriven detonation expands outwardly, the probe (i.e., already in the post-detonation area) temperature and pressure decrease gradually towards a constant value, whereas the HRR is reduced to almost zero.

Figure 11(a) shows the evolutions of the reactive hotspot along the monitoring line from 10-110 nanoseconds in case F ($H_2$+$O_2$ hotspot, $p_s$ = 150$p_0$). The reader should be reminded that due to one-dimensional nature of early shock/detonation structures, the results in Fig. 11 do not exhibit azimuthal variations. Homogeneous isochoric reactions occur inside the spot, leading to quickly increased pressure and temperature. The HRR peaks at about 41.5 nanoseconds, and then quickly decreases to low values at around 110 nanoseconds. Meanwhile, the hotspot pressure and temperature remain unchanged from 90 to 110 nanoseconds, indicating the completion of chemical reactions in the hotspot. During this period, the premixture outside the hotspot remain intact; see Fig. 11(a). The hotspot reaction leads to increased pressure and temperature gradients at the hotspot vicinity, which plays an important role in initiating a detonation (Gu, Emerson & Bradley 2003).

Plotted in Fig. 11(b) are the state evolutions at the hotspot vicinity after 0.2 μs. Apparently, an outwardly propagating SF (the shock pressure is around 40$p_0$) emanated from the hotspot vicinity can be observed. It arrives at the probe at around 0.5 μs, resulting in a pronounced pressure rise, as shown in Fig. 10. A RF trails behind the shock, burning the compressed $H_2$+air mixture, featured by high HRR. Their mutual reinforcement quickly initiates a developing detonation, as found from the subsequent instants (0.7-1.1 μs) in Fig. 11(b). Furthermore, the corresponding evolutions of $\lambda_{CEM}$, a chemical explosive mode (Lu *et al.* 2010), along the monitoring line is shown in Fig 14(a). Finite $\lambda_{CEM}$ can be seen near the hotspot vicinity, manifesting the locally strong reactivity, which induces the immediate onset of detonation. Besides, there is a secondary inwardly propagating RF in the spot in Fig. 11(b). This may be attributed to chemical reactions as the local temperature drops due to thermal diffusion (see Fig. 11b). In addition, the hotspot evolution in case E is generally like that in F, but the shock arrives at the probe around 0.2 μs later, due to slower speed.

Differently, for non-reactive hotspots in case A-D, we can see from Fig. 10 that, although the maximum pressure exceeds the von Neumann values, their maximum temperatures are much lower than the respective von Neumann values. Furthermore, their peak HRRs that occur downstream of the hotspot are almost four orders of magnitude smaller than those in E and F. All these indicate that only shock compression occurs there and detonation has not developed yet. As shown in Section 3, detonations are ultimately initiated in case A and B. Therefore, it is interesting to further investigate how they are generated with a shock from the hotspot.

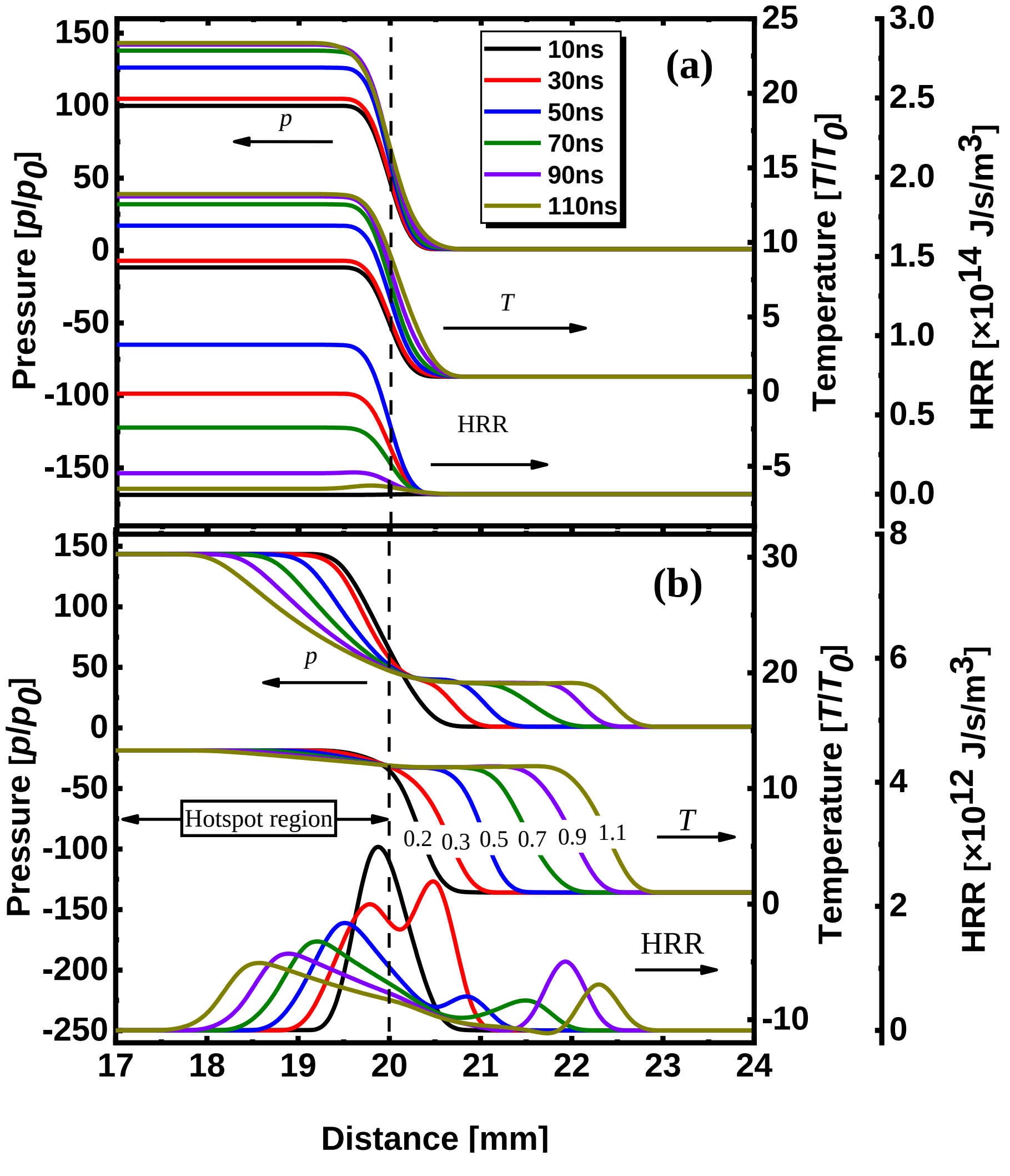

Figure 11: Changes of pressure, temperature, and HRR in the hotspot at (a) 10-110 nanoseconds and (b) 0.2-1.1 microseconds in case F. Time stamps in (b) are in microsecond.

We first look at case A, and the corresponding hotspot evolutions along the monitoring line from 0.8-2 μs are shown in Fig. 12. Note that the pressure inside the spot is maintained at the initial value, i.e., $250p_0$, since no reactions happen therein (not displayed in Fig. 12). At larger radii, e.g., at $R$ = 21-28 mm, the pressure peak first deceases and then increases, see the inset of Fig. 12. This is because the leading shock, generated at the hotspot vicinity, decays as it travels outwardly (see the HRR profiles, 0.8-1.4 μs). The peak pressure at 1.4 μs is around $22p_0$, 1.4 times the von Neumann value, whilst the peak HRR reaches around $1.5 \times 10^9$ J/m$^3$/s (not shown in Fig. 12). Subsequently, the reactions start in the shocked mixture (2-3 mm off the hotspot) at approximately 2 μs, and the peak HRR increases to around $5 \times 10^{11}$ J/m$^3$/s at 3-4 μs. This indicates the formation of shock-induced auto-igniting RF and its acceleration behind the leading SF (Gu, Emerson & Bradley 2003). As the RF couples to the leading SF, the detonation is initiated. In this case, $\lambda_{CEM}$ keeps zero close to the hotspot (see Fig. 14b), and increases immediately behind the SF and peaks at the RF from 1-2 μs. As the autoignition wave approaches the SF, the distribution of $\lambda_{CEM}$ shows double peaks at 3-4 μs and the $\lambda_{CEM}$ rises dramatically behind the SF. This detonation initiation fashion differs from those in case E and F, where the unburned mixture is directly ignited by the detonation from the hotspot. As such, it can be expected that detonation initiation beyond the hotspot is more affected by reactivity of the detonable mixture, besides the hotspot itself. The hotspot evolution in case B is like that in A, but with a delayed detonation initiation. The shock wave arrives at the probe 0.02 μs later than that in Case A, see Fig. 10.

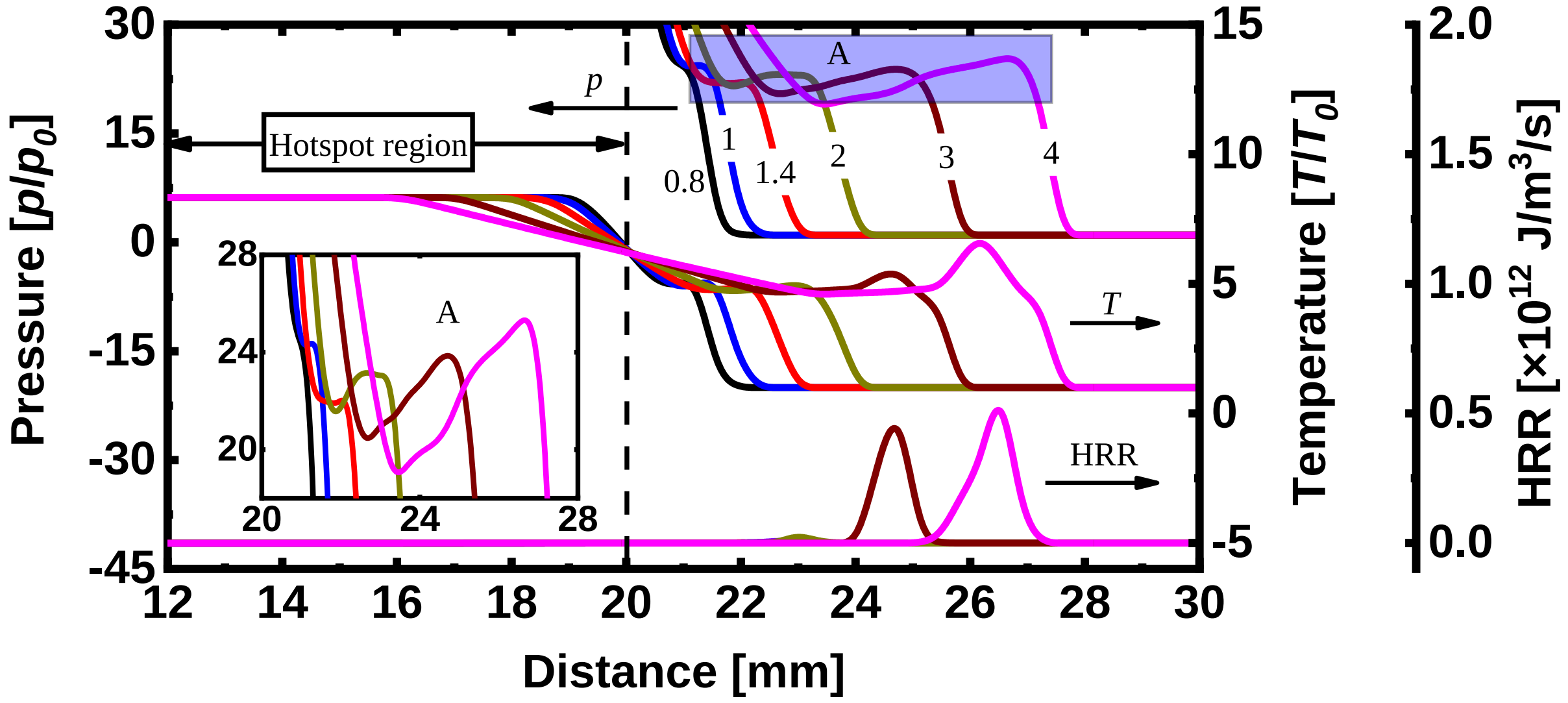

Figure 12: Changes of pressure, temperature, and HRR near the hotspot at 0.8-2 μs in case A. Time stamps in microsecond.

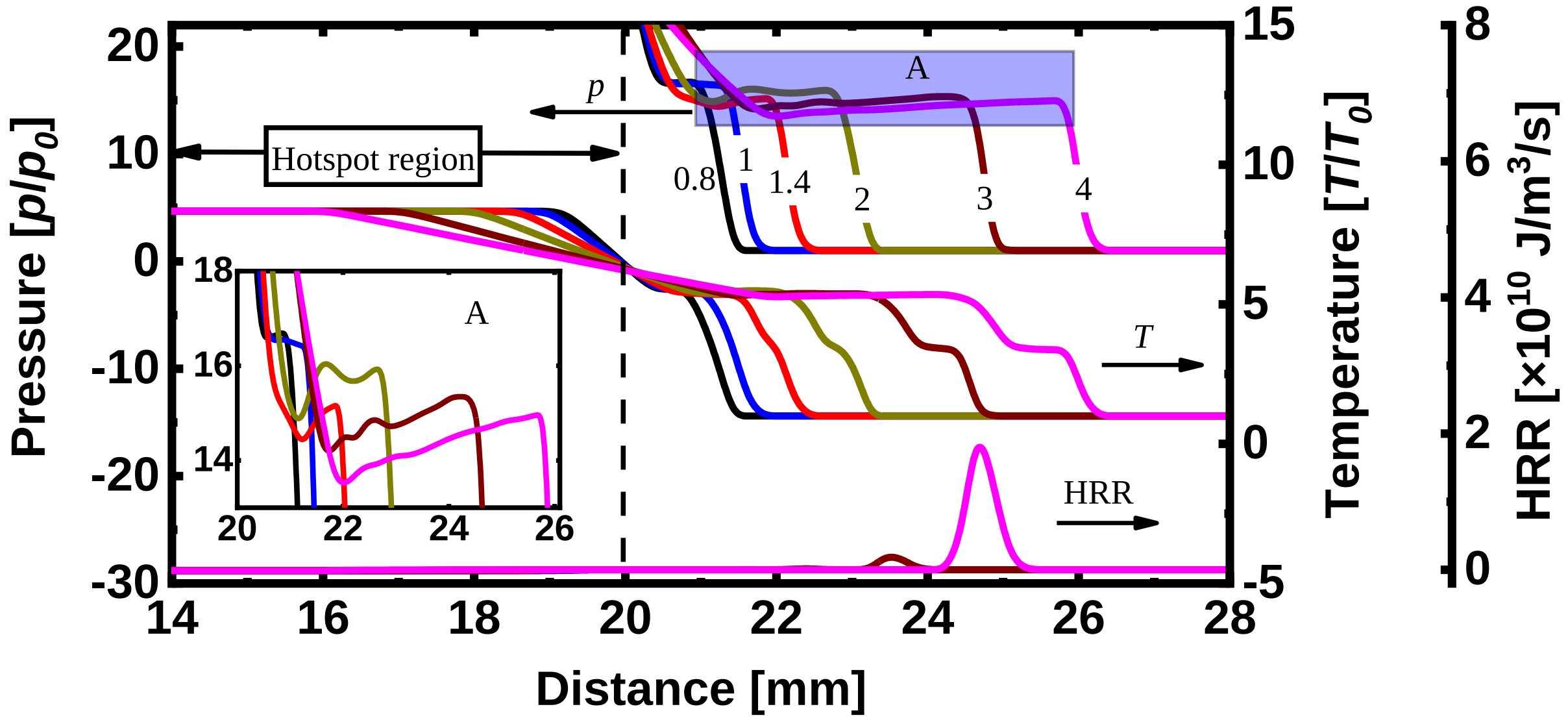

Figure 13: Changes of pressure, temperature, and HRR in the hotspot at 0.8-4 μs in case D. Time stamps in microsecond.

Figure 13 shows the hotspot evolution along the monitoring line in case D. Similar to case A, the pressure peak drops initially (0.8-1.4 μs) and then increase (2μs). However, it drops again from 2 to 4 μs; see the inset in Fig. 13. This is because although the reaction contributes to the shock amplification, the peak pressure keeps only 14-16$p_0$, close to the von Neumann value (15.1$p_0$) (Shepherd 2021). Furthermore, the thermally neutral zone is continuously lengthened from 2 to 4 μs. The auto-ignition RF is also ignited like case A. Nonetheless, the RF cannot synchronize with the leading SF, eventually failing to initiate the detonation. Different from case A, the distribution of CEM shows one single peak at 3-4 μs in case D (see Fig. 14c), and an obvious plateau appears during the rise of the $\lambda_{CEM}$ (see A region in Fig. 14c) at 4 μs due to the decoupling of RF and SF. Actually, during the later propagation of the shock, no detonation development (e.g., deflagration-to-detonation transition) is found. In another detonation failure case, C, hotspot evolution is generally similar to that in case D.

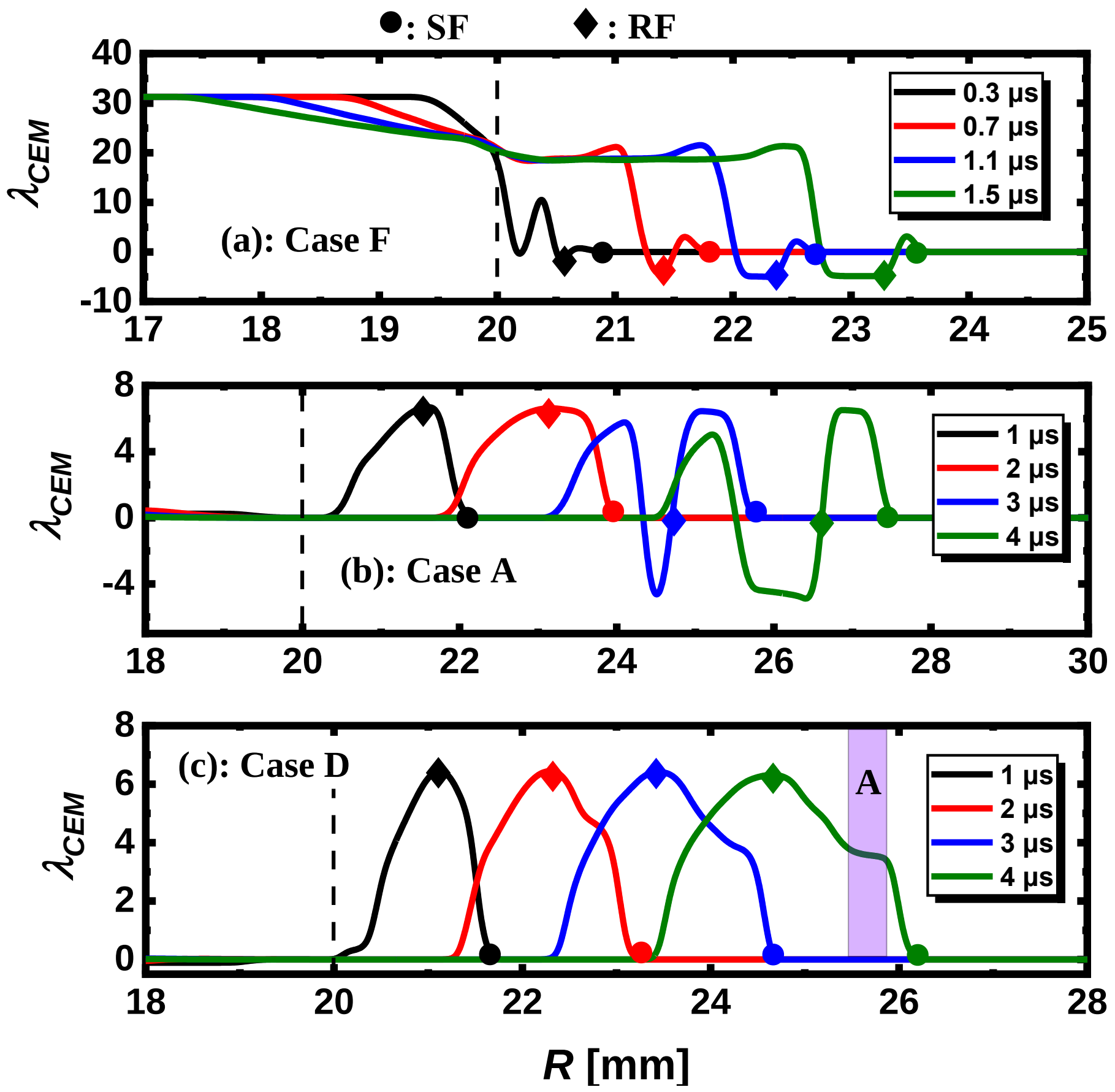

Figure 14: Evolutions of the chemical explosion mode along the monitoring line during the hotspot development. (a) case F, (b) case A, (c) case D. Dotted lines: hotspot vicinity.

To generalize the hotspot effects, we conduct a series of simulations by changing the hotspot pressure for both non-reactive (i.e., air) and reactive (i.e., $H_2$+$O_2$ and $H_2$+air) hotspots. Figure 15 shows the probe pressure ($p_p$) and temperature ($T_p$) as functions of the hotspot pressure ($p_s$) at the probe ($R$ = 21 mm). Since the isochoric reactions first take place inside the spot for reactive hotspots (i.e., $H_2$+$O_2$, $H_2$+air), we also present the maximum pressure ($p_{max}/p_0$) and temperature ($T_{max}/T_0$) when the reactions are completed in the hotspot.

Generally, detonations are not initiated directly from all non-reactive hotspots regardless of its pressure; see the probe temperature in Fig. 15(a). When the hotspot pressure is $p_s$ = 50-300$p_0$, the detonation can be initiated somewhere beyond the hotspot only when $p_s \geq 200p_0$. Moreover, detonation initiation depends on whether the shock is strong enough to ignite the detonable mixture followed. For the studied cases, the peak probe pressure should be at least 1.5 times the von Neumann pressure for successful initiation. It is seen from Fig. 15(a) that the probe pressure $p_p$ increases monotonically with $p_s$, but with a decreasing slope, whilst the probe temperature $T_p$ increases almost linearly with $p_s$. The detonation cannot be initiated beyond the hotspot when $p_s$ decreases to $\leqslant 150p_0$. Although the probe pressure for $p_s = 100p_0$ or $150p_0$ slightly exceeds the von Neumann spike, the RF is too weak and eventually the detonation initiation is not successful (marked as "Fail" in Fig. 15a).

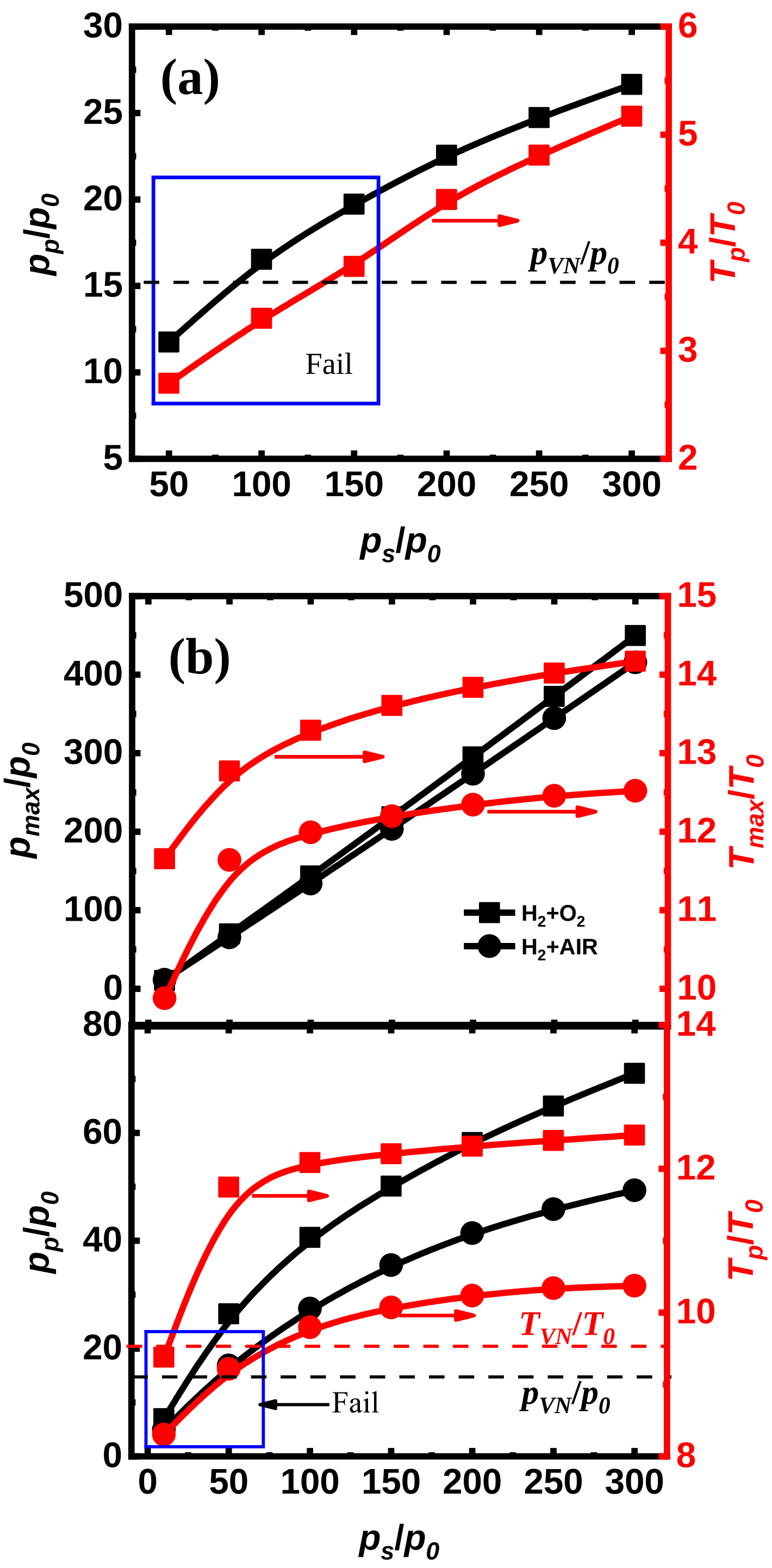

Figure 15: Change of pressure and temperature at the probe ($R$ = 21 mm) with different hotspot pressures: (a) non-reactive hotspot and (b) reactive hotspot.

For the reactive hotspot cases in Fig. 15(b), the detonation can be directly initiated due to the significant gradient of thermochemical states at the spot vicinity under appropriate $p_s$ (i.e., $p_s \geqslant 50p_0$ for $H_2$+$O_2$ hotspot and $p_s \geqslant 100p_0$ for $H_2$+air hotspot). No detonations can be initiated when the hotspot pressure decreases to $10p_0$ for the $H_2$+$O_2$ hotspots and $\leqslant 50p_0$ for the $H_2$+air hotspots (annotated with "Fail" in Fig. 15b). Based on the current simulations, detonation initiation by the shock beyond the reactive hotspot is not observed.

As the hotspot pressure increases, the peak hotspot pressure $p_{max}$ due to the isochoric reactions increases linearly (see Fig. 15b). However, the peak hotspot temperature increases dramatically when $p_s \leqslant 150p_0$; beyond that, it grows slowly when the hotspot pressure further increases. This is because the chemical equilibrium moves towards the exothermic reaction direction and gradually approaches the limit. Furthermore, both probe pressure $p_p$ and temperature $T_p$ monotonically increase with the initial pressure of the reactive hotspots, and the growth rate decreases with the hotspot pressure. This is similar to what is seen from the non-reactive hotspots in Fig. 14(a).

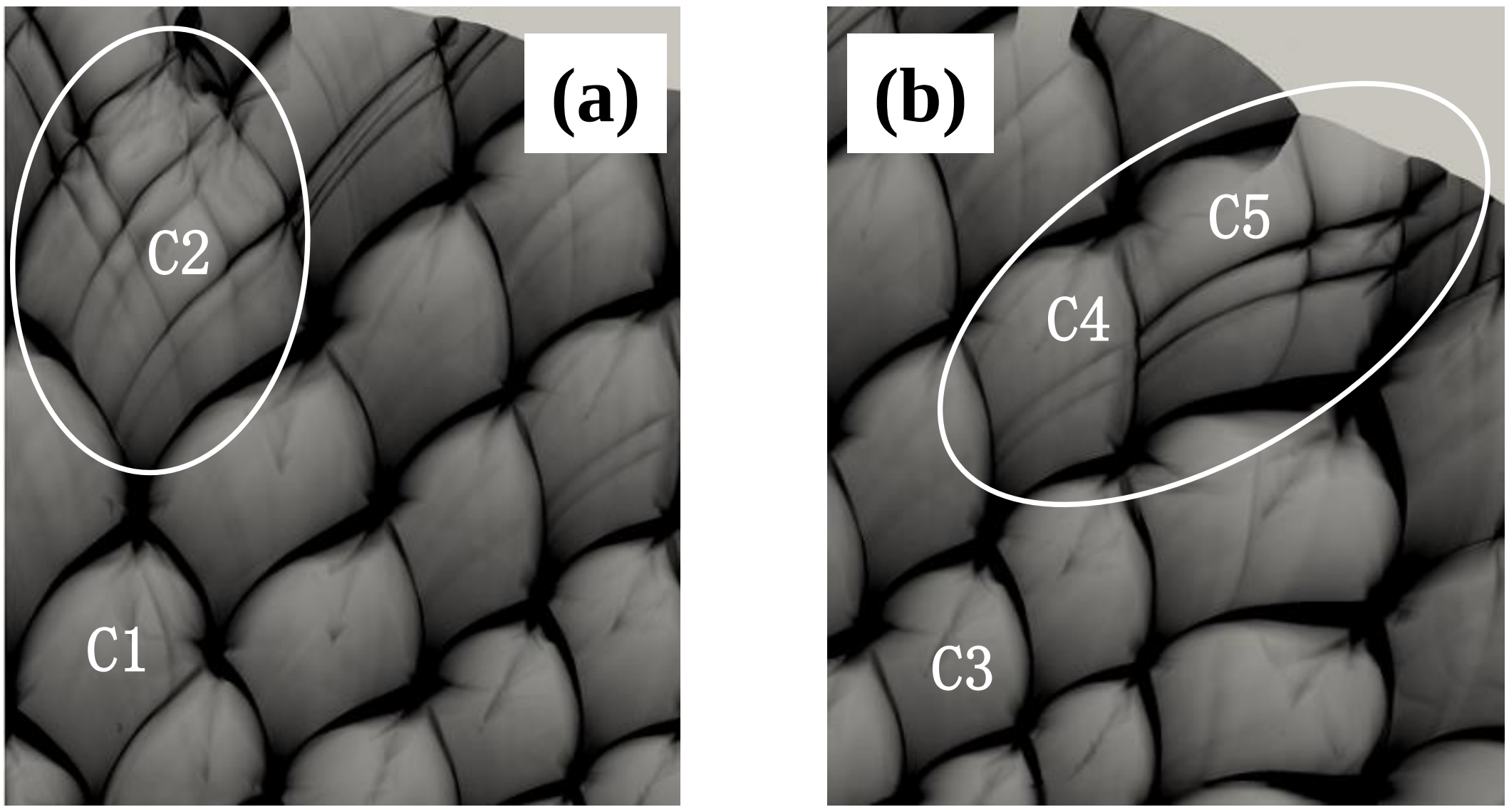

Figure 16: Detonation cell diverging in case B: (a) abrupt pattern and (b) gradual pattern.

*4.2. Detonation cell diverging and coalescence*

As the cellular instability increases to a certain threshold value as the detonation propagates outwardly, additional transverse waves would be generated to match the growing surface of the detonation for self-sustaining propagation (Han *et al.* 2017; Jiang *et al.* 2009). In this section, we will discuss two patterns of cell diverging observed from case B (see Fig. 3b): abrupt and gradual diverging, respectively, in Figs. 16(a) and 16(b). In the abrupt pattern, as the cell C1 grows and develops into C2 after the next triple point is generated, the secondary peak pressure is elevated in C2, signifying the formation of new cells in it. Nonetheless, for the gradual pattern, relatively weak pressure waves are generated in C4. With the shock interaction between C4 and C5, the forgoing weak pressure become stronger in C5, and eventually new secondary cells are generated in C5. The growth rate of cell size from C1 to C2 is 50%, much higher than that (22%) from C3 to C4. This difference is responsible for various cell diverging patterns and the details will be presented in Figs. 17 and 18.

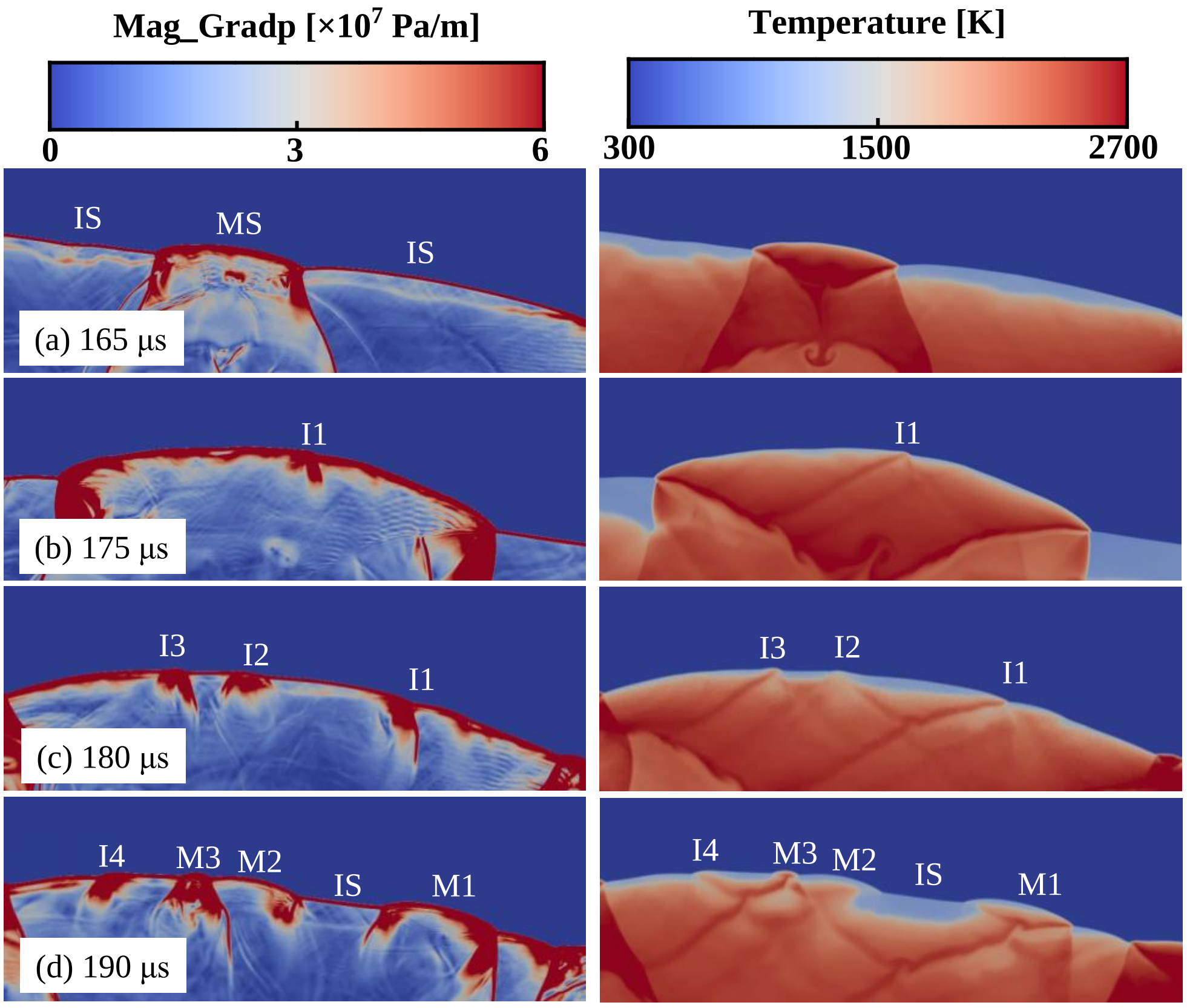

Figure 17: Evolution of abrupt cell diverging: pressure gradient magnitude (left column) and temperature (right column).

Figure 17 shows the pressure gradient magnitude and temperature distributions during the abrupt diverging transient. The Mach stem, MS, from the triple point is smooth initially (e.g., at 165 μs). Its velocity is higher than the adjacent incident shocks, IS. As the curved MS propagates outwardly, its surface significantly increases, leading to decreasing number of cells per unit area of the detonation front. Consequently, instability 1 (I1 in Fig. 17b) occurs along the Mach stem, leading to front wrinkling (Shen & Parsani 2017). Moreover, the temperature near I1 is higher than that of the surrounding (see 175 μs), corresponding to higher local reactivity. The shock near I1 propagates with a greater speed, causing a convex front. The transverse wave originating from I1 propagates circumferentially and interact with IS, further intensifying I1, see 180 μs. Other instabilities, e.g., I2 and I3, appear due to the similar mechanism. As the MS decays, a thin gap is generated between the leading shock and reaction front, and the temperature therein is lower (see 180 μs), indicating the increased induction time. The mixtures near the instabilities are more explosive, eventually resulting in a RF. The RF then couples to the SF, and new Mash stems are developed, e.g., at 190 μs. MS1-3 originates from I1-3, respectively. Meanwhile, new instability I4 is generated due to increased detonation surface. The detonation front is divided into several sections, with staggered MS and IS.

Evolution of the gradual diverging pattern is detailed in Fig. 18. The occurrence of instabilities is like those in the abrupt pattern, but they are too weak to induce new cells when the detonation expands. Instead, the instability amplification during the interactions between two adjacent shocks plays an important role in the diverging process. In Fig. 18, the instabilities I1 and I2 only occur along the IS at 176 μs. As the adjacent MS expands, it collides with the neighbouring shock. I1 is obviously intensified after collision with the transverse wave of the MS; see 183 μs. This also induces the wrinkled Mach stem. Besides, the interaction between two opposite transverse waves originating from I2 and MS, respectively, can be observed, as marked by green circle in Fig. 18(b). After the instability amplification from IS to the adjacent MS, the subsequent diverging behaviour is like that from the abrupt pattern. At 195 μs, new hotspots evolve from the instability, generating new Mach stems. Finally, new cellular features, including MS, IS and TW, appear at 206 μs.

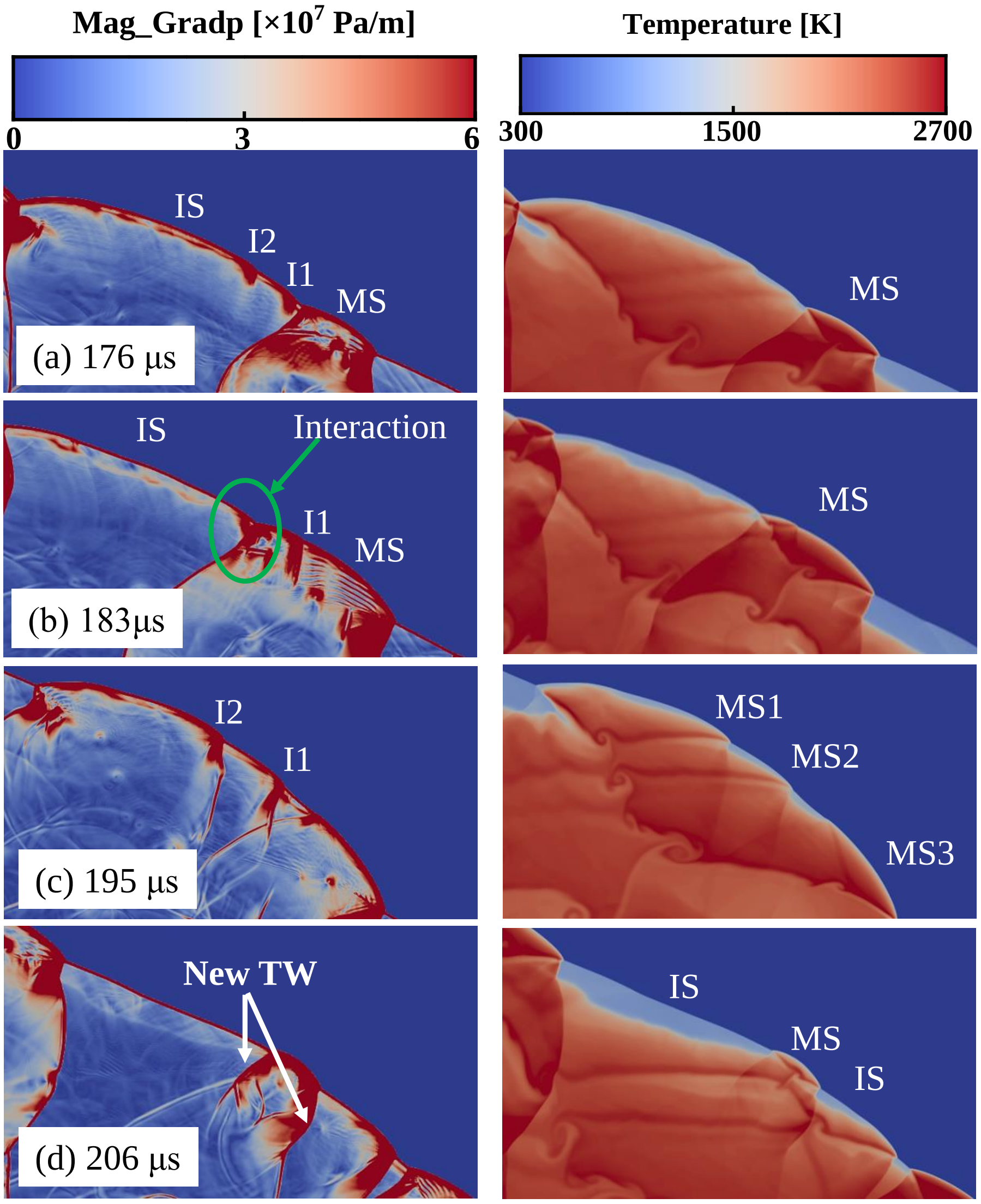

Figure 18: Evolution of gradual cell diverging: pressure gradient magnitude (left column) and temperature (right column).

Another phenomenon worthy of discussion at the cell-growing stage is cell coalescence. In contrast to the diverging behaviours, the cell coalescence always takes place where the initial irregular cell pattern dominates, which subsequently generates larger local cell. Figure 19 shows a detailed coalescence process in case E, as demonstrated in Fig. 5(a). To better illustrate the underpinning mechanism, a relative cell size (i.e., cell thickness in this work) perpendicular to the cell family direction is introduced, denoted by $l_a$-$l_c$ in Fig. 19. The cell thickness of C2 is only around a third of C1 or C3, giving a high instability as the detonation propagates outwardly. Furthermore, as the C1 and C3 grow along the cell family, C2 further decreases until ultimate disappearance.

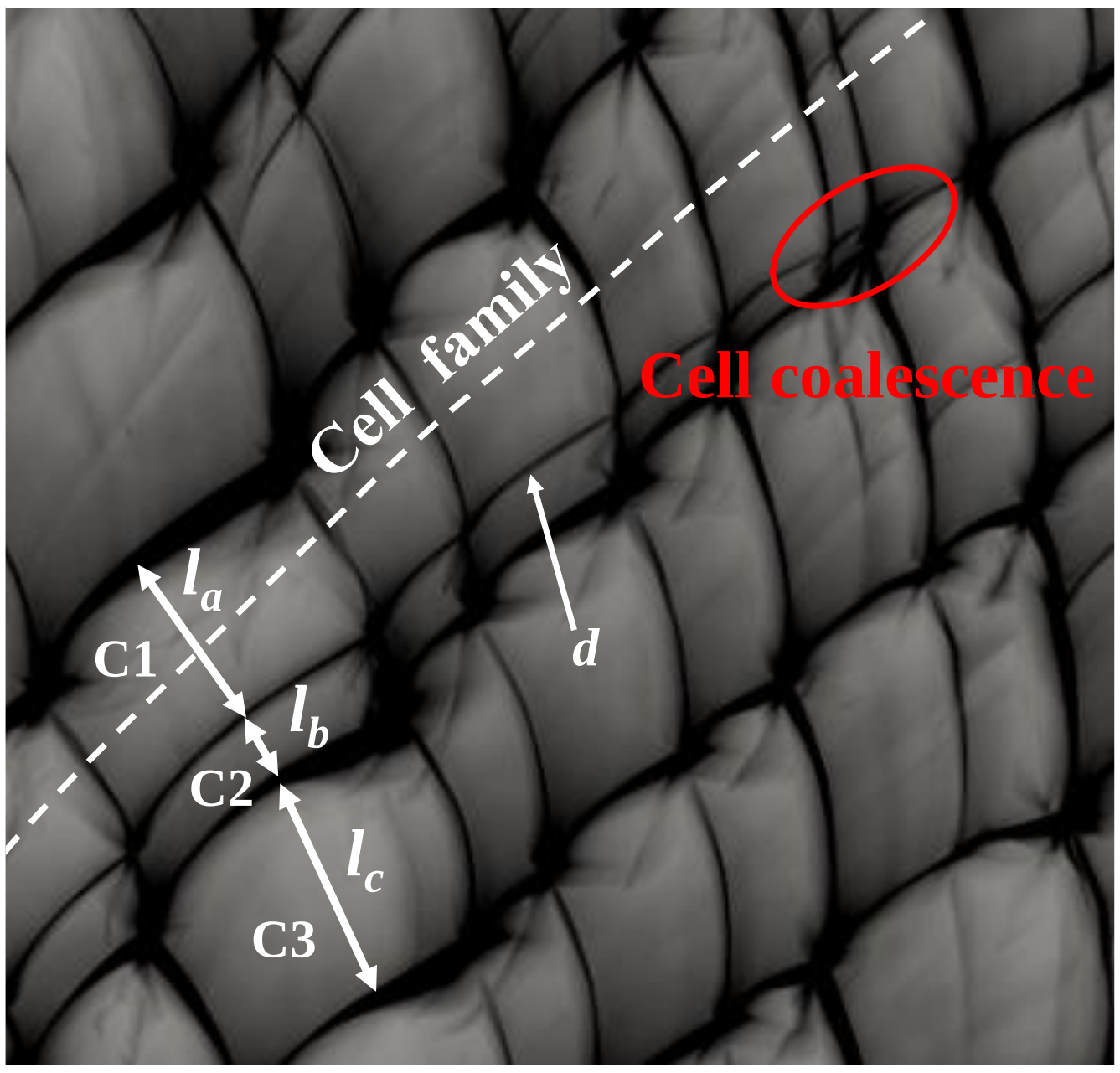

Figure 19: Distributions of the peak pressure trajectories in a cell coalescence process in Fig. 5(a).

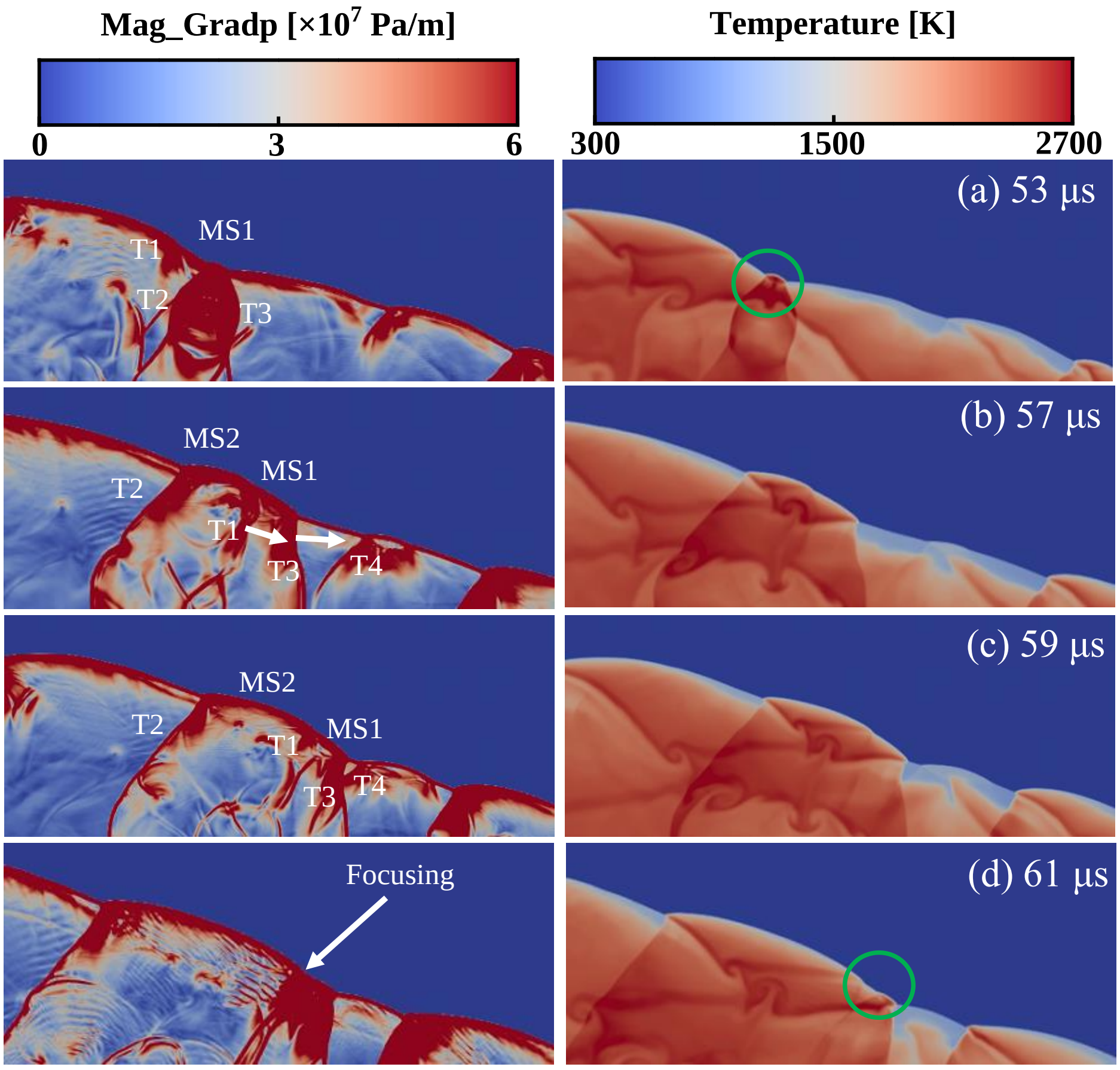

Figure 20: Evolution of the cell coalescence in Fig. 19: pressure gradient magnitude (left column) and temperature (right column).

Figure 20 shows the cell coalescence transient, using the pressure gradient magnitude and temperature distributions at successive instants. At 53 μs, MS1 is generated with intense chemical reactions accompanied by two strong transverse waves, T2 and T3. Meanwhile, a relatively weak transverse wave T1 propagates towards T2. After the collision between T1 and T2, MS2 is generated near MS1, and its speed is larger than that of MS1. As such, two adjacent MS's (MS1 and MS2) form a large MS along with three TWs, among which T1 and T3 propagate at the same direction. The trajectory of T1 can be found in Fig. 20 (see *d*). As the detonation evolves, T1 is gradually approaching T3. As T1 coalesces with T3 and further collides with T4 (see 59-61 μs, Fig. 19), a new hotspot is generated with local high reactivity (see the circled region, 61 μs), accompanied by the disappearance of T3.

To quantify the cell variations as the detonation wave propagates outwardly, in Fig. 21 we show the calculated cell size from uniform and non-uniform detonable mixtures. Here the cell size is approximated by the radial distance between A and B along the arc $R_1$, as shown in Fig. 21(a). The fitted line is plotted by the proportional relation between a certain arc and the number of cell-family, as shown in Figs. 21(b) and 21(c). The cell-family number remains constant before diverging in case A, B, and F (see Figs. 3 and 5b), whilst decreases due to cell coalescence for case E and G-I. (see Figs. 5a and 7). Consequently, the fitted lines feature constant for case A, B, F and increasing slope for case E and G-I, respectively, see Fig. 21(b) and 21(c). Note that they are obtained from the cell-growing stage in all cases. The theoretical cell size with the Ng correlation (Ng, Ju & Lee 2007) and the experimental data by Stamps & Tieszen (1991) are also presented for reference in the uniform case.

Generally, the cell growth rate decreases (hence cell-family number increases) with hotspot pressure for non-reactive hotspots A and B. This can be ascribed to higher detonation strength in case A as the cellular detonation initially forms. Besides, a lower cell growth rate of case F can be found from Fig. 21 (b) due to higher overdrive.

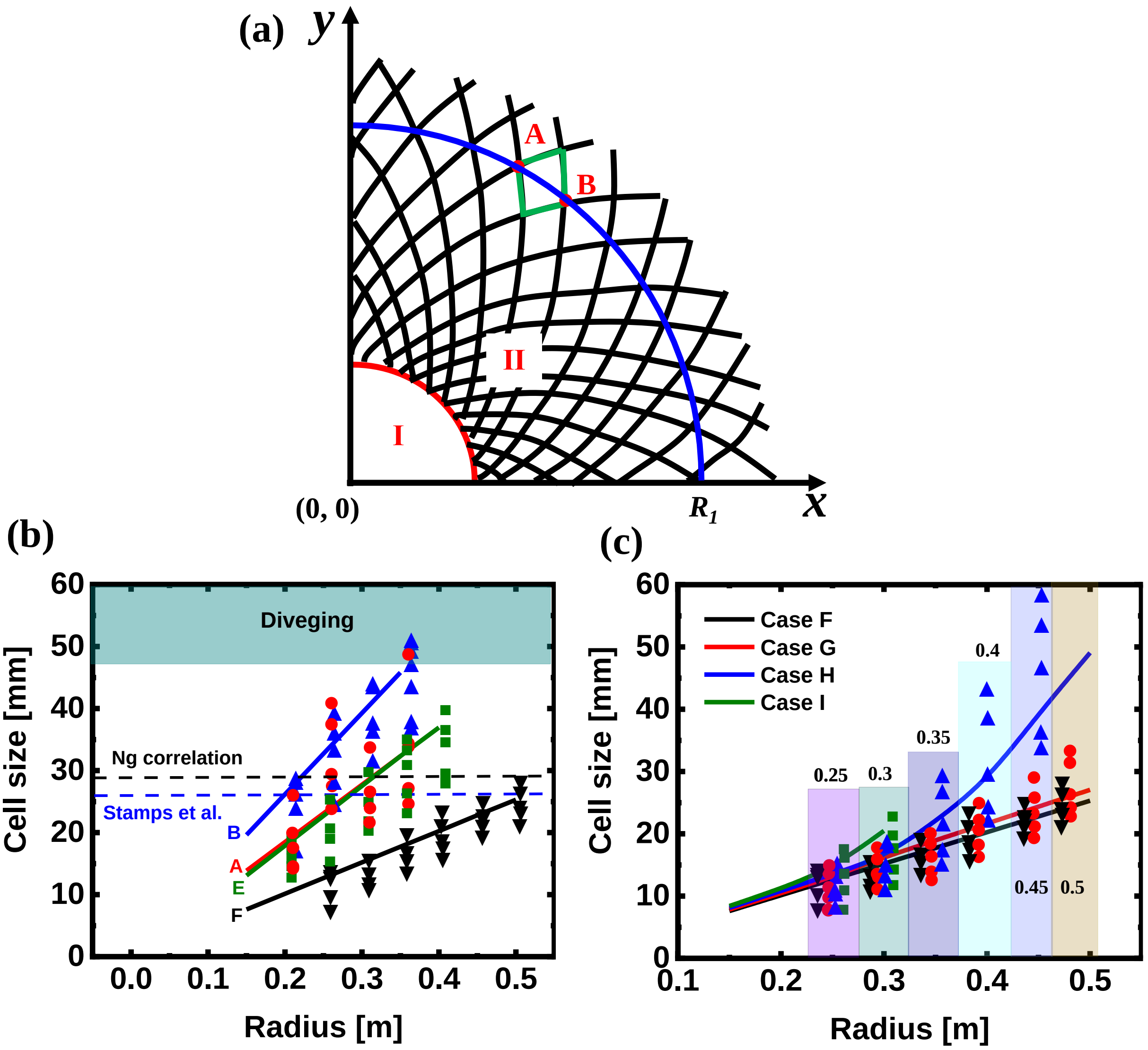

Figure 21: Detonation cell size at different hotspot properties: (a) schematic of cell estimation; (b) change of detonation cell size along the radial distance in cases A, B, E, and F; (c) change of detonation cell size along the radial distance in cases F-I. Black dashed line: theoretical data from Ng correlation (Ng, Ju & Lee 2007). Blue dashed line: experimental data with initial pressure of 0.26 atm (Stamps & Tieszen 1991). Number above the histogram: radius of the arc where the cell samples are obtained.

For cases A, B and E, the detonation cell starts to diverge when the cell size approaches a certain value. It can be inferred from Fig. 21(b) that this threshold is a value greater than the corresponding theoretical and experimental cell sizes, under which condition the detonation propagates at a relatively stable state (Lee 1984). When the cylindrical detonation expands, the collisions between adjacent transverse waves become difficult due to increased spacing. Accordingly, the cellular instability significantly increases, especially as the average cell size continuously grows beyond the characteristic cell size. This enhanced instability leads to generation of new transverse waves to sustain detonation propagation (Jiang *et al.* 2009). In the studied cases, the threshold value is 1.4-2 times the characteristic cell size under the same mixture condition. It is worth noting that real three-dimension detonation structure is very complex involving irregularity and inhomogeneity of the detonation cell (Pintgen *et al.* 2003; Crane *et al.* 2022), which may lead to greater fluctuations of the threshold than that in the current simulations.

For case F, the sizes of cell samples at various radii are closer to the fitted line due to the more uniform distribution. Furthermore, the maximum cell size is well below the corresponding theoretical and experimental values, and thus it remains at the cell growing stage even at larger radius, see Fig. 5(b).

In contrast to the uniform mixture cases, only cell coalescence happens in the nonuniform cases. In these cases, the cell size dramatically increases as the ER decreases to match the increased half-reaction length, which makes the detonation more unstable. For better illustration, we put the cell samples horizontally adjacent for different cases at the same radii, as annotated by the columns in Fig. 21(c). Generally, the detonation cell evolutions are similar between case F ($\varphi$ = 1) and G ($\varphi$ = 1→0.9) due to close mixture reactivities. For case H ($\varphi$: 1→0.5), the cell growth rate increases considerably across 0.3 m. Meanwhile, the cell sizes become more scattering when $R$ = 0.3-0.5 m, due to cell coalescences, see Fig. 7(c). For case I ($\varphi$: 1→0), the maximum cell size reaches about 22 mm at $R$ =0.3 m (corresponding to $\varphi$ = 0.42), where the detonation extinction happens.

### *4.3. Hydrodynamic structure*

We will further discuss the hydrodynamic thickness variations in expanding cylindrical detonations. The hydrodynamic thickness is the distance between the sonic plane and the shock front (Lee & Radulescu 2005). In case B, a self-sustaining diverging detonation is generated. Figure 22 shows the time sequence of shock-frame Mach number in this case. They are from four stages, i.e., overdrive (5-45 μs), cell formation (54-78 μs), growing (100-170 μs), and diverging stages (195-238 μs). At 5 μs, an overdriven detonation propagates outwardly from the hotspot, and behind it a sub-sonic region exists. This subsonic region is further elongated radially at 20-45 μs, which makes the detonation more susceptible to rarefaction effects from its behind. At 20 μs, the flow field behind the detonation is separated into three regions, marked by *a*, *b*, and *c*. Specifically, in region *a*, the dissipation of the hotspot happens, and the gradually reduced *Ma* is ascribed to the increasing flow speed. Regions *b* and *c* are the burned zone of the deflagration and detonation, respectively. All three regions increase as the detonation expands. Consequently, the transverse disturbance, which is intensified by the curvature, renders the detonation cellularized at 45 μs (Han *et al.* 2017).

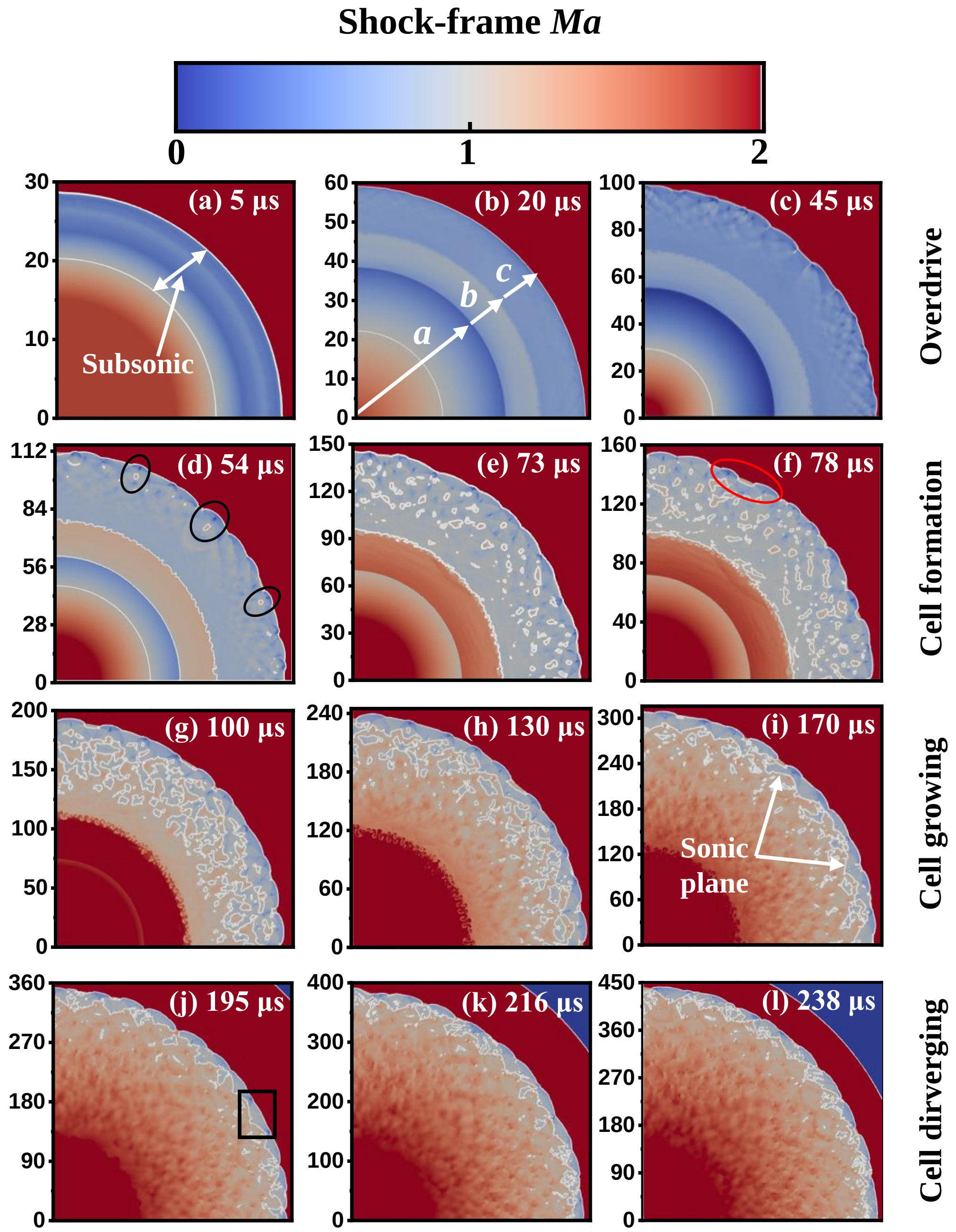

Figure 22: Changes of shock-frame Mach number in case B. Axis label in millimeter. White line: $Ma$ = 1 isolines.

At 54 μs, the detonation cellularization becomes more pronounced, accompanied by some scattered supersonic pockets behind the detonation wave. The effect of the expansion waves on detonation front varies at different circumferential positions. Since the expansion waves cannot penetrate the supersonic pocket to attenuate the detonation (Weber & Olivier 2003), the local detonation fronts ahead of the supersonic pockets show higher speed, as shown in the circled regions. Meanwhile, there is an increased Mach number behind the leading shock due to the decreased flow speed at 54-78 μs. Especially, a supersonic ring is generated from the subsonic region *b* with two extra sonic lines. At 78 μs, more small supersonic zones are generated behind the detonation front, indicating the enhancement of the hydrodynamic fluctuations. This fluctuation applied to the expansion wave further influences the local detonation intensity, and eventually promotes the formation of the triple-point structure of the detonation front; see the red circles at 78 μs.

At the cell growing stage (100-170 μs), the localized supersonic pockets gradually coalesce with each other, and are extended to the entire subsonic zone. At 170 μs, a relatively clear sonic region appears immediately behind the SF. Furthermore, the hydrodynamic thickness becomes constant over the time, indicating the formation of the freely propagating cylindrical detonation. It is reported in Ref. (Radulescu *et al.* 2007) that the mechanical and thermal fluctuations decays from a large magnitude (6-10%) close to the shock front to a negligible intensity (0.5-1%) at the sonic surface. Therefore, the disturbance behind the sonic plane has little effects on the detonation front, and the detonation evolution is only governed by the available energy release and product expansion between the sonic plane and leading shock (Lee & Radulescu 2005).

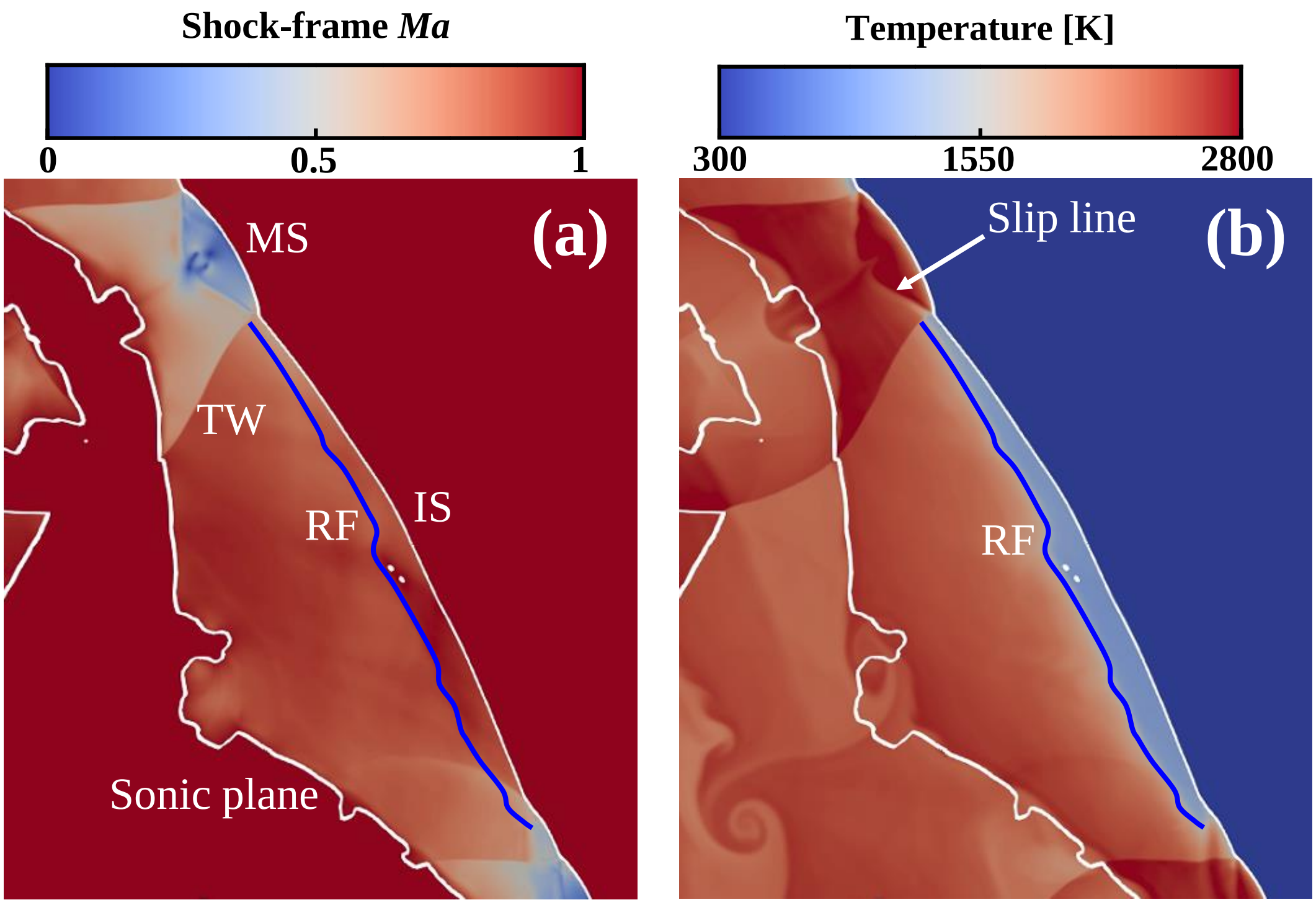

Figure 23: Contours of (a) shock-frame Mach number and (b) temperature from the box in Fig. 22(j). White line: *Ma* = 1. Blue line: RF.

At the cell diverging stage (195-238 μs), the subsonic zone only appears between the leading shock and sonic plane as a "sawtooth" pattern. This is due to the combined effects of transverse wave, incident shock and Mach stem (Radulescu *et al.* 2007). Figures 23(a) and 23(b) shows the enlarged Mach number contour (the *Ma* range adjusted to 0-1.2) and the temperature distribution from the box in Fig. 22(j). In the subsonic zone, the Mach number is nearly constant behind the incident shock, except for the induction zone where the Mach number is higher due to lower temperature, as shown in Fig. 23(b). However, behind the Mach stem, the Mach number increases towards the sonic plane. This is because rapid reactions take place immediately after the Mach stem, which dramatically raises the local temperature and flow speed. A distinct *Ma* discontinuity exists at the interface of post-MS and post-IS products due to the transverse waves. Owning to the different expansion of post-MS and post-IS products, the sonic plane shows convex behind MS, whilst concave behind IS, which leads to the "sawtooth" pattern of the sonic plane.

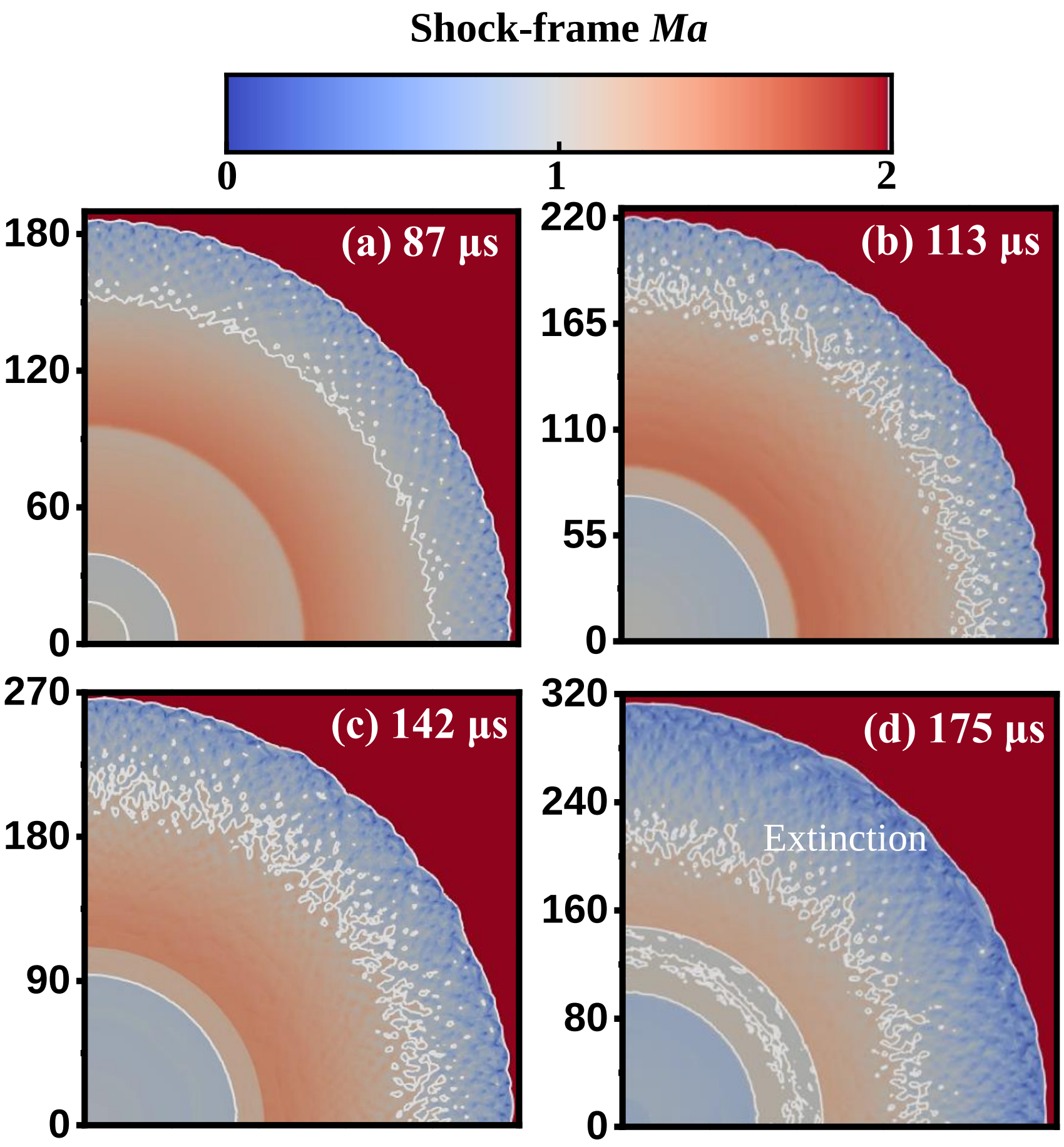

Figure 24: Changes of shock-frame Mach number with times in case I. Axis label in millimeter. White line: $Ma$ = 1 isolines.

Figure 24 shows the time sequence of the Mach number in case I, in which the detonation is ultimately quenched as the equivalence ratio approaches zero. Overall, the shock Mach number gradually decreases when it runs outwardly. At 87 μs, an overdriven detonation with small cells is generated. As a result, the subsonic region behind the leading shock is relatively long (about 30 mm). From 113-142 μs, plenty of supersonic pockets occur in the subsonic region just like 73-78 μs in Fig. 22. However, generation and coalescing of the supersonic spot do not make the subsonic region shrink; instead, the subsonic region is further elongated due to the weakened shock intensity at 142 μs. Further downstream the detonation decays to an inert shock with relatively smooth front at 175 μs.

## 5. Conclusions

Two-dimensional cylindrical detonation direct initiation in hydrogen/air mixtures are computationally studied. The effects of hotspot property and mixture composition gradient on detonation initiation are studied. The main conclusions are summarized as below.

(1) For nonreactive hotspot, initiation fails for low hotspot pressure ($p_s$ = 100$p_0$ or 150$p_0$) and critical regime dominates for high hotspot pressure ($p_s$ = 200$p_0$ or 250$p_0$) in which three stages occurs, including no cell, growing cell and diverging cell. Supercritical regime dominates for both reactive hotspot ($H_2$+$O_2$ or H2+air, $p_s$ = 100$p_0$). Detonation is directly initiated from the reactive hotspot, whilst it is initiated somewhere beyond the nonreactive hotspot through the coupling of the leading shock and reaction front.

(2) The detonation cell size increases almost linearly with the radius at the cell-growing stage, which implies that the cell-family number of the cellular detonation determines the growth rate of cell size in a cylindrical detonation. Furthermore, cell diverging only occurs as the local cell size exceeds the characteristic cell size of a certain value.

(3) Two cell diverging patterns are identified, i.e., abrupt and gradual patterns, respectively. The abrupt diverging is attributed to the generation and intensification of the instability as the cell grows, whilst the gradual diverging is mainly caused by the instability amplification during the interactions between two adjacent shocks. Besides, the cell coalescence occurs if much irregular cells initially form and the cell with smaller cell thickness merges to the bigger one as the detonation expands. As such, the cell-family number is reduced and the local cells grow faster, which leads to an earlier diverging behavior.

(4) As the mixture ER decreases linearly from unity at the hotspot vicinity to a certain value at $R$ = 0.5 m, the detonation experiences self-sustained propagation, highly unstable propagation (with local extinction), and global extinction with ER: 1→0.9, 1→0.5 and 1→0, respectively. In particular, highly unstable detonation arises from multiple cell coalescences, and detonation extinction occurs where the induction time is highly lengthened and unburned pockets occur.

(5) Hydrodynamic structure analysis is also conducted for both uniform and nonuniform mixtures. For a self-sustaining detonation case (air hotspot, $p_s$ = 100$p_0$), the hydrodynamic thickness first increases at the overdrive stage, then decreases as the detonation cells are generated, and eventually reaches almost a constant at the cell diverging stage in which the sonic plane exhibits a "sawtooth" pattern. This is ascribed to the different expansion of post-MS and post-IS products. For detonation extinction case (φ: 1→0), hydrodynamic thickness continuously increases from the overdriven state to extinction and no "sawtooth" sonic plane occurs since no self-sustaining detonation is generated.

## Acknowledgement

This work used the computational resources of the National Supercomputing Centre, Singapore (https://www.nscc.sg/). XJ is supported by The China Scholarship Council.